\documentclass[3p]{elsarticle}

\usepackage{hyperref}
\usepackage{subcaption}

\usepackage{amsmath}
\usepackage{multirow}
\usepackage{graphicx}
\usepackage{float}
\usepackage{xcolor}
\usepackage[linesnumbered,ruled,vlined]{algorithm2e}
\usepackage{adjustbox}
\RestyleAlgo{ruled}
\SetKwComment{Comment}{/* }{ */}

\journal{Vehicular Communications}

\begin{document}

\begin{frontmatter}

\title{Evaluation and improvement of ETSI ITS Contention-Based Forwarding (CBF) of warning messages in highway scenarios}

\tnotetext[mytitlenote]{\textbf{Published as: Oscar Amador, Manuel Urue\~na, Maria Calderon, Ignacio Soto, Evaluation and improvement of ETSI ITS Contention-Based Forwarding (CBF) of warning messages in highway scenarios, Vehicular Communications, Volume 34, 2022, 100454. The final version of record is available at \href{https://doi.org/10.1016/j.vehcom.2022.100454}{ https://doi.org/10.1016/j.vehcom.2022.100454.}}}
\tnotetext[mytitlenote]{This work was partially supported by the Agencia Estatal de Investigaci\'on (AEI, Spain) through the ACHILLES project (PID2019-104207RB-I00/AEI/10.13039/501100011033) and by the Madrid Government (Comunidad de Madrid-Spain) under the Multiannual Agreement with UC3M in the line of Excellence of University Professors (EPUC3M21), and in the context of the V PRICIT (Regional Programme of Research and Technological Innovation).}
\tnotetext[titlenote2]{We gratefully acknowledge support from the Swedish Knowledge Foundation (KKS) for the "Safety of Connected Intelligent Vehicles in Smart Cities -- SafeSmart" project (2019--2023) and from the ELLIIT Strategic Research Network.}

\author[Halmstad]{Oscar Amador}
\ead{oscar.molina@hh.se}

\author[UNIR]{Manuel Urue\~na}
\ead{manuel.uruena@unir.net}

\author[UC3M]{Maria Calderon}
\ead{maria@it.uc3m.es}

\author[UC3M]{Ignacio Soto\corref{mycorrespondingauthor}}
\cortext[mycorrespondingauthor]{Corresponding author, current address: Departamento de Ingenier\'{\i}a de Sistemas Telem\'aticos, Universidad Polit\'ecnica de Madrid, 28040 Madrid (Madrid), Spain.}
\ead{ignacio.soto@upm.es}

\address[Halmstad]{School of Information Technology; Halmstad University; Halmstad 30118; Sweden}
\address[UNIR]{Escuela Superior de Ingenieros y Tecnolog\'{\i}a, Universidad Internacional de la Rioja; 26006 Logro\~no; Spain}
\address[UC3M]{Departamento de Ingenier\'{\i}a Telem\'atica; Universidad Carlos III de Madrid; 28911 Legan\'es (Madrid); Spain}

\begin{abstract}
This paper evaluates the performance of the ETSI Contention-Based Forwarding (CBF) GeoNetworking protocol for distributing warning messages in highway scenarios, including its interaction with the Decentralized Congestion Control (DCC) mechanism. Several shortcomings of the standard ETSI CBF algorithm are identified, and we propose different solutions to these problems, which are able to reduce the number of transmissions by an order of magnitude, while reducing the message end-to-end delay and providing a reliability close to 100\% in a large area of interest.
\end{abstract}

\begin{keyword}
ETSI Intelligent Transport Systems (ITS) \sep Decentralized Environmental Notification Message (DENM) \sep Contention-Based Forwarding (CBF) \sep Decentralized Congestion Control (DCC) \sep Duplicate Packet Detection (DPD)
\end{keyword}

\end{frontmatter}

\section{Introduction}

Intelligent Transportation Systems (ITS) aim to improve road safety and traffic efficiency by exchanging messages among vehicles and with roadside stations. In particular, the dissemination of warning messages is used by ITS applications to warn drivers of potential hazards such as an accident, roadworks or bad road conditions. The vehicle that detects the specific incident initiates the transmission of the warning message, which is retransmitted in multi-hop mode to reach an entire area of interest (i.e., the area affected by the incident).

The \textcolor{black}{European Telecommunications Standards Institute (ETSI)} has specified several sets of messages for ITS applications. In particular, Decentralized Event Notification Messages (DENMs)~\cite{etsiDEN} are event-triggered messages used to alert users of hazardous events. DENMs are disseminated in the area affected by the event using the services of the ETSI GeoNetworking protocol. The ETSI GeoNetworking protocol~\cite{etsiNewGeoNetworking} enables multi-hop GeoBroadcast communication where a message is broadcast over a geographical area in a multi-hop basis (i.e., vehicles rebroadcast the message until the whole area of interest is covered). The ETSI GeoNetworking protocol specifies Contention Based Forwarding (CBF)~\cite{etsiNewGeoNetworking} as a forwarding strategy to perform multi-hop GeoBroadcast, being the default option when the potential forwarder is inside the area of interest.

CBF is a receiver-based forwarding algorithm (i.e., for each incoming packet the vehicle decides whether to retransmit the packet) based on timers.  Packets are transmitted using the broadcast capability of the wireless access technology, and the receiving neighbours include the incoming packet in a CBF packet buffer (if it was not already there), associated with a contention timer. The receiving node (i.e., a candidate forwarder) calculates this timer duration based on the relative position of the ego node with the previous sender (i.e., the node that initially generated the DENM or the previous forwarder), such that the neighbour farthest away (i.e., the best forwarder) will set the timer to a lower value than vehicles closer to the previous sender. Therefore, it is expected that the first timer to expire will be that of the best forwarder, which will retransmit the packet, and once the rest of the receiving neighbours (i.e., other candidate forwarders) receive this new transmission, they will cancel their own retransmissions by removing the packet from their CBF buffers.

ETSI has developed a set of specifications to enable communication of CBF messages over different access technologies in the dedicated 5.9 GHz frequency range, where IEEE 802.11p/ETSI ITS-G5~\cite{etsiMediaDependentG5} is a possible technology. \textcolor{black}{It is well-known} that vehicle density and the data load generated by applications can lead to network congestion in the radio medium, causing packet collisions and potentially unlimited medium access delay. Thus, the ETSI ITS specifications include a Distributed Congestion Control (DCC) mechanism \cite{etsiNewDcc} to be used with the ITS-G5 access layer, which ensures the radio medium operates in an efficient regime. This DCC function at the access layer regulates the message sending rate at each transmitter by keeping the waiting packets stored at DCC queues according to its priority (Traffic Class). The latest version of the DCC specification~\cite{etsiNewDcc} includes an adaptive variant based on an algorithm called LIMERIC~\cite{Limeric2013}, which uses a linear control system that compares the measured channel load with a target value and adapts the message sending rate so that medium occupancy converges to the target value.

A major challenge for the ETSI CBF forwarding mechanism, when used as multi-hop GeoBroadcast protocol, is to control the network overhead (number of times the same message is transmitted), while maximizing its reachability (to reach as many vehicles as possible in the area of interest). In fact, the deployment of the full ETSI ITS stack with the CBF forwarding protocol operating over DCC (with queues that may add delay) has undesirable effects on the operation of the CBF protocol, and jeopardizes the goal of keeping the network overhead low and achieving high reachability in the area of interest.

In this paper, we analyze the behavior of the CBF algorithm specified by ETSI~\cite{etsiNewGeoNetworking} to disseminate DENMs over an area of interest in highway scenarios. We identify several limitations, with focus on the effect of the DCC mechanism for ITS-G5 on ETSI CBF performance. The contributions of our paper are the following:
\begin{enumerate}
    \item Identification of different situations in which the ETSI CBF mechanisms cause large network overloads. We show that its current design generates waves of retransmissions of the same packet in the area of interest.
    \item Identification of different situations that lead ETSI CBF to cancel the forwarding of a packet before reaching the border of the area of interest.
    \item Analysis of the impact of ETSI DCC on the two previously-identified phenomena. DCC limits the ability of CBF to cancel the transmission of duplicate packets (i.e., a duplicated packet cannot be cancelled if it is waiting in the DCC queue); and DCC reorders transmitters with respect to what CBF expects, meaning that worse transmitters can cancel retransmission on better forwarders that would improve packet progress in the area of interest.  
    \item We propose several improvements to ETSI CBF to solve the above-mentioned issues, which yields lower transmission overhead and better reachability in a large area of interest.
\end{enumerate}

The remainder of the paper is organised as follows: section~\ref{subsec:background} provides some background information. Specifically, it introduces the ETSI standardization for Intelligent Transport Systems (ITS) and presents the previous work on multi-hop GeoBroadcast. Section~\ref{subsec:contributions} identifies several deficiencies of the ETSI CBF protocol operating over DCC and proposes multiple mechanisms to solve them, which are the main contributions of this paper. Next, section~\ref{sec:evaluation} defines the simulation-based evaluation scenario of ETSI CBF and the proposed improvements, and analyzes their performance results. Finally, section~\ref{subsec:conclusions-fw} draws conclusions and presents future work.

\section{Background}
\label{subsec:background}

\subsection{ETSI ITS Architecture}
\label{subsec:etsi-its-architecture}

\begin{figure}[t!]
	\centering
	\includegraphics[width=\textwidth]{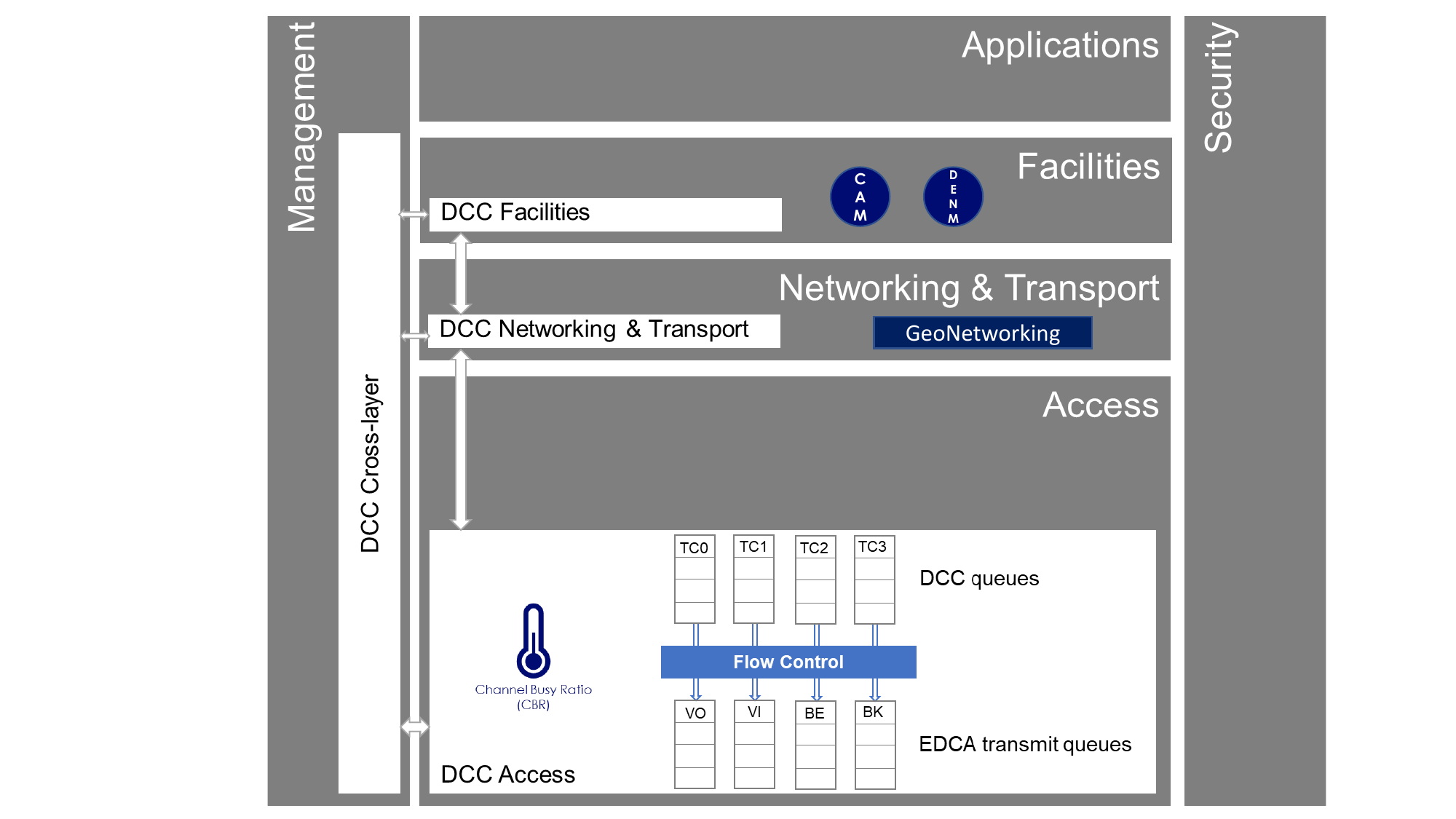}
	\caption{ETSI ITS Architecture}
	\label{fig:ETSI_Arch_1}
\end{figure}

ETSI has developed a set of specifications to enable vehicle-to-vehicle communication in the 5.9 GHz band, which define the ETSI Intelligent Transport System (ITS) stack. The ETSI ITS architecture is shown in Fig.~\ref{fig:ETSI_Arch_1}, with the ETSI GeoNetworking~\cite{etsiNewGeoNetworking} protocol located at the Networking\&Transport layer. This protocol can work over different access technologies for short-range wireless communications in the 5.9 GHz band, such as IEEE 802.11p/ETSI ITS-G5~\cite{etsiMediaDependentG5} or LTE-V2X~\cite{etsiMediaDependentLTE}.

ETSI GeoNetworking is a network protocol that makes use of geographical positions to route a packet towards its destination. ETSI GeoNetworking allows several kinds of communications, namely: GeoUnicast (i.e., the destination is an individual node), Single-hop Broadcast (i.e., all nodes at one hop), topologically-scoped Broadcast (i.e., all nodes in a n-hop neighbourhood), GeoAnycast  (i.e., the destination is any node inside an area), as well as GeoBroadcast. GeoBroadcast allows the dissemination of packets over geographical areas (i.e., all the nodes inside the area) and hence its interest for the deployment of safety applications (i.e., warning all vehicles in an area upstream and/or downstream of a hazardous incident).

At the Facilities layer, Cooperative Awareness (CA)~\cite{etsiCA} and Decentralized Environmental Notification (DEN)~\cite{etsiDEN} services are the cornerstones of ETSI \textcolor{black}{Cooperative ITS (C-ITS)}. These basic services support a great variety of ITS applications. Cooperative Awareness service makes use of Cooperative Awareness Messages (CAMs) which contain information regarding the status (e.g., position and heading) of the generating node. These CAMs are sent periodically to all nodes at one hop (Single-hop Broadcast). On the other hand, \textcolor{black}{DENMs} enable DEN services and are the basis for Road Hazard Warning (RHW) applications~\cite{etsiRHS}, \textcolor{black}{identified as one of the "day 1" ITS applications~\cite{C-ITS2016, C-ROADS} to be deployed in real scenarios}. RHW applications use DENMs to alert road users of a variety of events (e.g., an animal on the road, an accident, roadworks, or a stationary vehicle). DENMs are sent by the source node that detects the dangerous situation, and are disseminated over a geographical area (i.e., circular, rectangular or ellipsoidal) as multi-hop GeoBroadcast messages (see Fig.~\ref{fig:ETSI_geobroadcast}). 

To provide differentiated quality of service to ITS applications, GeoNetworking protocol handles traffic classes to prioritize the network traffic. Thus, each packet has an associated Traffic Class (TC) that indicates its requirements on data transport. There are four different priorities: TC0 to TC3, with TC0 being the highest one. CAMs correspond to traffic class TC2. In the case of a DENM at the source node (i.e., the vehicle that originates the DENM), this packet corresponds to Traffic Classes TC0 or TC1 (highest priorities), but when forwarded it is assigned a traffic class TC3 (lowest priority).

\subsubsection{ETSI Decentralized Congestion Control}
\label{subsec:dcc}

The ETSI ITS specifications include a \textcolor{black}{DCC} mechanism~\cite{etsiNewDcc}. Congestion control is essential in vehicular networks to ensure that the radio medium operates at an efficient regime. In particular, ETSI ITS architecture includes DCC entities located at several layers (see Fig.~\ref{fig:ETSI_Arch_1}) with the DCC Cross-layer management plane enabling the communication between DCC entities at different layers.

As for DCC at the Access Layer, the ETSI DCC mechanism specification includes, in its latest version~\cite{etsiNewDcc}, an adaptive variant to be used with the ITS-G5 access layer~\cite{access:2020}. This adaptive variant is based on an algorithm called LIMERIC \cite{Limeric2013}, which performs congestion control by independently regulating the message sending rate at each transmitter (Transmit Rate Control). For this purpose, it employs a linear control system using the channel occupancy as input. The DCC Access entity senses the Channel Busy Ratio (CBR), i.e., the percentage of time the channel is busy, which is a metric of the channel occupation. This approach aims at collectively converging to a predefined channel utilization level, i.e., a target Channel Busy Ratio (CBR). Packets waiting at the access layer due to the rate limitation are stored in DCC queues according to their traffic class (see Fig.~\ref{fig:ETSI_Arch_1}). DCC Access Layer (DCC gatekeeper) dequeues packets according to this allowed sending rate. After a packet is transmitted, the time that the node waits before the next packet is dequeued (i.e., \textit{$t_{go}$}) is updated following the adaptive approach to a time between 25 and 1,000 milliseconds.

At the Facilities Layer, message-generation frequency can be also modulated according to the rate DCC assigns to the service. This is the case of Cooperative Awareness basic service~\cite{etsiCA}, where CAM frequency is dynamically adapted according to vehicle dynamics and channel occupation.

\subsubsection{ETSI ITS Contention-Based Forwarding}
\label{subsec:cbf}

\begin{figure}[t]
	\centering
	\includegraphics[width=\textwidth]{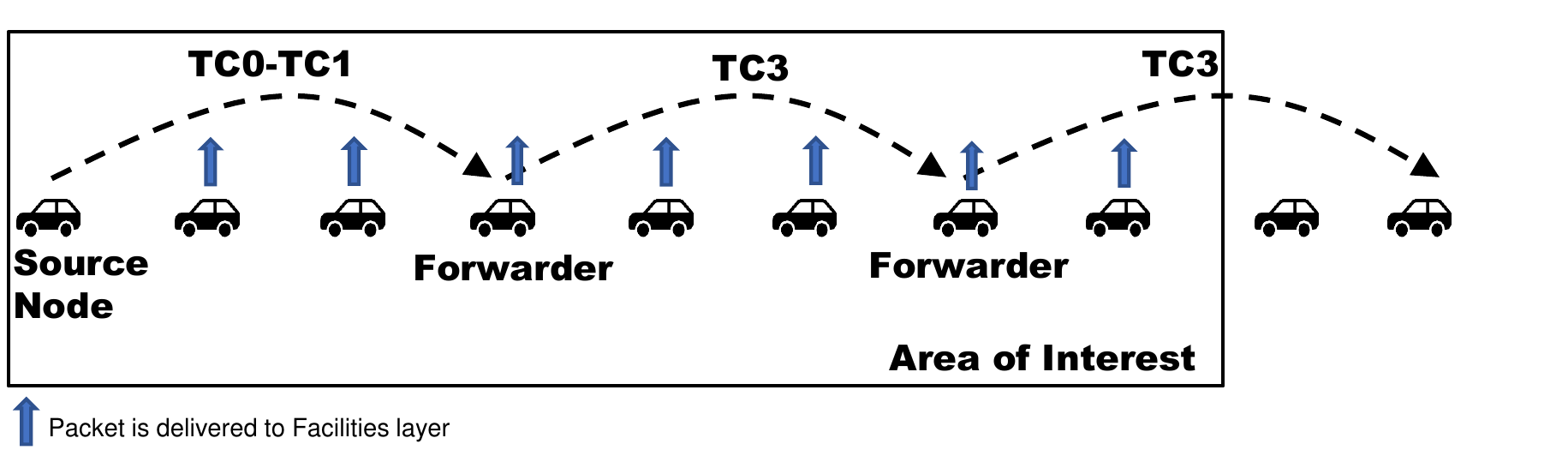}
	\caption{Multi-hop GeoBroadcast}
	\label{fig:ETSI_geobroadcast}
\end{figure}

The ETSI GeoNetworking protocol specifies two basic forwarding strategies to perform multi-hop GeoBroadcast when forwarders are located inside the area of interest (see Fig.~\ref{fig:ETSI_geobroadcast}). These are: Simple Area Forwarding algorithm and \textcolor{black}{CBF}. The specification also includes several non-area forwarding strategies, such as Greedy Forwarding \cite{tomatis:2015}, but in this paper we will focus on area-forwarding algorithms.

The Simple Area Forwarding algorithm is a plain flooding mechanism i.e., all nodes rebroadcast a packet as soon as they receive it. The algorithm includes a strategy to avoid duplicates (DPD - Duplicate Packet Detection) and limit flooding based on the source node address, the packet sequence number, a maximum hop limit, and a maximum lifetime.

The \textcolor{black}{CBF} algorithm is an opportunistic mechanism. In CBF, when a packet is broadcast (i.e., packets are transmitted using the broadcast capability of the wireless access technology), the receiving neighbours insert the packet into a CBF packet buffer (if it was not already there) and start a timer that is inversely proportional to the distance to the sender (i.e., previous forwarder or source node). Thus, the neighbour farthest away from the sender will have the smallest timer and it will try to forward the packet first. This timer (\textit{$T_{CBF}$}) is calculated according to the following equation~\ref{eq:TempCBF}:

\begin{equation}
\label{eq:TempCBF}
T_{CBF} = \begin{cases}
T_{CBF-MAX - \frac{T_{CBF-MAX} - T_{CBF-MIN}} {DIST_{max}} \times DIST }&   \text {if $DIST$ $\leq$ $DIST_{MAX}$} ,\\
T_{CBF-MIN} &   \text {if $DIST$ $>$ $DIST_{MAX}$}
\end{cases}
\end{equation}

where \textit{$T_{CBF-MAX}$} is the maximum time the packet is stored at the CBF buffer (default value 100\,ms), \textit{$T_{CBF-MIN}$} is the minimum time the packet is stored at the CBF buffer (default value 1\,ms), \textit{$DIST_{MAX}$} is the theoretical maximum communication range of the wireless access technology (default value 1000\,m), and \textit{$DIST$} is the distance between the forwarder that has received the packet and the sender.

When $T_{CBF}$ expires, the node re-broadcasts the packet. Theoretically, this results in an efficient use of the channel, as the distance being covered with each forwarding step is maximised. The node farthest away from the previous sender has the smallest $T_{CBF}$ and it is the first to forward the packet. Then, the algorithm needs to prevent the other nodes that received the packet (from the previous sender) from re-broadcasting it. To achieve this, if a node receives a packet that is already in its CBF packet buffer, this node drops the received packet and also removes the copy from its CBF packet buffer. 

\textcolor{black}{The ETSI GeoNetworking~\cite{etsiNewGeoNetworking} specification also includes an area advanced forwarding algorithm that has not been considered in this work. This area advance forwarding algorithm includes mechanisms to (1) combine greedy forwarding and CBF, (2) choose potential CBF forwarders only from those outside a sector defined by the greedy transmission, and (3) allow the retransmission of the same packet several times in the same forwarding area.}

\subsection{Related Work}
\label{subsec:related-work}

Multi-hop broadcast has been a research topic for many years. The simplest approach to this problem is by using flooding, where each node in the area of interest forwards an incoming message once. It is well-known that flooding makes an inefficient use of network resources, but the main problem is that it leads to the broadcast storm problem. A comprehensive survey on broadcast storms caused by flooding can be found in~\cite{tseng:2002}. Many research papers have been proposed to alleviate the broadcast storm issue and to make an efficient use of network resources. A common approach has been to choose a subset of forwarders to alleviate network storms and reduce unnecessary retransmissions, i.e., duplicated packets.

The various approaches for next-hop forwarders selection can be classified into two categories, according to where the forwarding decision is made: at the sender, i.e., each forwarder decides the next-hop forwarder(s)~\textcolor{black}{\cite{Torrent2007}}\cite{,Sahoo:2011,bai:2009}; or at the receiver node, i.e., for each incoming message the forwarder decides whether or not to retransmit it. This last approach has received quite attention in the case of Vehicular Area Networks (VANETs), since it does not require vehicles to know their neighbour positions which implies less network overhead. The election of the next forwarders can be probabilistic~\cite{zeng:2018, Abbasi:2020, REINA:2015}, in which receiving nodes decide to forward an incoming message according to certain probability value. Probabilistic techniques exhibit good performance in network latency, but high overhead, i.e., multiple forwarders might decide to forward the same message, resulting in collisions and inefficient channel use. An interesting alternative, preferred by many research works, is using contention (i.e., delay-based or timer-based approaches)~\cite{DDT:2000,briesemeister:2000}, in which all nodes receiving an incoming packet become candidate forwarders and contend with each other for message forwarding. The contention time is based on a forwarding delay that is computed according to different parameters (e.g., positions, vehicle speed, vehicle density). The contender node with the minimum timer duration transmits the message first and other nodes cancel their own scheduled transmissions of the same message once they find out that the message has been already forwarded by another vehicle (contention-based protocols have to use the broadcast address as destination link address).

The timer-based techniques have been extensively studied for unicast communications (i.e., at the application layer, the destination is an individual node)~\cite{lima:2009,li:2008,chen:2007}, and more recently, they have also been rapidly gaining ground for multi-hop broadcast (i.e., the message is rebroadcast repeatedly until the packet is disseminated over the entire area of interest)~\cite{Baiocchi:2016,lenardi:2007,chuang:2013}. A widely-studied aspect has been the combination of several parameters to set the contention timer~\cite{yang:2008, lenardi:2006}, and dynamically adjusting it to different conditions, such as vehicular density~\cite{chuang:2013}, speed~\cite{yang:2008}, heading~\cite{xu:2017} or physical Signal-to-Noise Ratio (SNR)~\cite{Abbasi:2020} values. The problems of waiting times~\cite{Rajendran:2013} and high end-to-end latency~\cite{yoo:2015,Bujari:2019,chuang:2013,Abbasi:2020} have been also of interest in the literature. Other works tackle the issue of reducing collisions (e.g., when multiple vehicles are located close to each other, and consequently having similar contention timers) that leads to poor performance~\cite{yoo:2015}. Similarly, \cite{Witt:2005} and \cite{lenardi:2007} attempt to reduce the number of unnecessary retransmissions, which also result in a low system performance. The applicability of timer-based approaches to vehicular sparse networks has also been the subject of research in the past, extending the coverage of the vehicular network with the support of road side infrastructure~\cite{Meijerink:2019}. Other papers address the die out problem that occurs when the multi-hop dissemination process is stopped due to forwarding decisions~\cite{REINA:2015}. An approach to minimize this die out issue consists of counting the number of copies for a given message, and if a candidate forwarder has not already seen a certain number of copies, it retransmits the message~\textcolor{black}{\cite{Torrent2007}} \cite{REINA:2015}.

\textcolor{black}{These multi-hop broadcast works do not consider a congestion control mechanism operating under the contention-based forwarding protocol. Interestingly, this is not the case of a relevant prior work~\cite{Torrent2007} where a contention-based strategy is combined with the selection of one specific forwarder (i.e., next-hop forwarder). The performance evaluation is carried out considering a power transmission control DCC mechanism operating underneath. However, this power transmission control DCC mechanism does not make use of a DCC queue as the ETSI DCC adaptive approach does, and which interaction with the ETSI forwarding protocol is studied in this paper.}

\textcolor{black}{CBF}~\cite{CBF:2003} is an opportunistic  timer-based technique, widely accepted~\cite{Kuhlmorgen:2015} by the scientific and industrial communities, in which the forwarding delay is computed according to the candidate forwarder relative position with the previous sender (i.e., the source or the previous forwarder). In this way, the message is spread further along the area of interest. ETSI ITS standardization specifies CBF as a forwarding solution for multi-hop GeoBroadcast dissemination~\cite{etsiNewGeoNetworking, tomatis:2015}. This algorithm is the default option when the potential forwarder is inside the area of interest. Some works in the past have evaluated the performance of ETSI ITS GeoNetworking~\cite{xu:2017, Kuhlmorgen:2015}, however only a few works~\cite{Kuhlmorgen:2016,Bellache:2017a, Kuhlmorgen2020} have considered the complete ETSI communications stack, including DCC. 

The work in \cite{Bellache:2017a} considers the use of a DCC mechanism at different layers, aligned with the ETSI ITS standards, but it omits DCC at Access Layer. The paper proposes to adjust the CBF forwarding parameters according to channel load conditions: the number of retransmissions and the maximum forwarding delay (i.e., the maximum time a packet is in the CBF buffer), i.e., under high channel load the number of retransmissions is reduced and the maximum forwarding delay is increased. 

The work in \cite{Kuhlmorgen2020} focuses on the issue of unnecessary retransmissions originated by DCC queues (i.e., a CBF packet cannot be cancelled once it is waiting at a DCC queue). However, their solution does not consider the related problem of choosing the best forwarder, i.e., delays in the DCC queues imply that the first node transmitting the packet may not be the best forwarder, topologically wise. This effect impairs CBF performance, as we will see later. 

The work in \cite{Paulin2015} identifies two relevant problems that are overlooked in other works on CBF: the need for a \textcolor{black}{DPD} mechanism that persists even if the message is no longer at the CBF buffer, and the possibility that two very close forwarders transmitting at the same time (without cancelling each other) results in the cancellation of all future transmitters, which stops the forwarding of the packet. For the second problem, \cite{Paulin2015} proposes two possible solutions: to use a randomized cancellation of packets in the CBF buffer, so that the arrival of a packet already in the buffer does not always result in its cancellation; or to introduce a progress check, so that a packet only cancels the retransmission if the new transmitter represents more progress towards the destination than itself, and if not, updates the CBF timer according to the new transmitter. However, it only studies the use of the CBF algorithm as a non-area forwarding mechanism (i.e., to reach a remote area of interest), it considers neither the use of a DCC mechanism, nor the delivery to the Application Layer, and the performance evaluation is carried out using only static vehicles.

\begin{algorithm}[p] \footnotesize
\caption{\textcolor{black}{ETSI GeoBroadcast forward and receiver operations~\cite{etsiNewGeoNetworking} (simplified)}}
\label{alg:ETSI_recv}
\KwIn{$p = \textrm{the received GeoBroadcast packet}$}
$SO$ is the source of $p$\;
$pos\_SO$ is the position vector of $SO$\;
$pos\_ego$ is the position vector of the ego vehicle\;
$RHL$ is the Remaining Hop Limit of $p$\;
$SCF$ if the Store-Carry-Forward bit of the Traffic Class field of $p$\;
$LocT$ is the Location Table\;
$LS$ is the Location Service\;
$BCBuffer$ is the broadcast forwarding packet buffer\;

ProcessBasicHeader($p$)\;
ProcessCommonHeader($p$)\;
Set $F \gets InsideAreaOfInterest(pos\_ego)$\;
\If{$F < 0 \textrm{ and } itsGnNonAreaForwardingAlgorithm = 1~(GREEDY)$}
{
  \tcc{Outside the area of interest}
  \If{$p$ in $dpl$}
  {
    Discard $p$\;
    Return\;
  }
}
\If{$F >= 0 \textrm{ and }itsGnAreaForwardingAlgorithm = 1~(SIMPLE)$}
{
  \tcc{Inside the area of interest}
  \If{$p$ in $dpl$}
  {
    Discard $p$\;
    Return\;
  }
}
DuplicateAddressDetection($P$)\;
\eIf{$LocT(SO) \textrm{ does not exist}$}
{
    $LocT(SO) \gets pos\_SO$\;
    $LocT(SO).IS\_NEIGHBOUR \gets FALSE$\;
    UpdatePacketDataRate($LocT(SO)$)\;
}
{
    $LocT(SO) \gets pos\_SO$\;
    UpdatePacketDataRate($LocT(SO)$)\;
}
\If{$F >= 0$}
{
    \tcc{Inside the area of interest}
    Indication($P.Payload$)\Comment*[r]{Pass the payload to upper protocol entity}
}
FlushPacketBuffers($SO$)\;
\eIf{$RHL <= 1$}
{
    Discard $p$\;
    Return\;
}
{
    $RHL \gets RHL - 1$\;
}
\If{$\forall LocT(SO).IS\_NEIGHBOUR = FALSE \in LocT \textrm{ and } SCF = TRUE$}
{
    \tcc{No neighbour exists}
    Add $p$ to $BCBuffer$\;
    Return\;
}
$ret \gets Forward(p)$
\If{$ret = 0\ (Buffered)$ or $ret = -1\ (Discarted)$}
{
    Return\;
}
MediaDependentProcedures($p$)\;
Transmit($p$)\;
\end{algorithm}

\begin{algorithm}[t!] \footnotesize
\caption{\textcolor{black}{ETSI Area Contention-Based Forwarding (CBF) algorithm~\cite{etsiNewGeoNetworking}}}
\label{alg:ETSI}
\KwIn{$p = \textrm{the GeoNetworking packet to be forwarded}$}
$pos\_ego$ is the ego position vector with Position Accuracy Indicator $acc\_ego$\;
$pos\_sender$ is the sender position vector in the Location Table with Position Accuracy Indicator $acc\_sender$\;
$buffer$ is the CBF packet buffer\;
$timeout$ is the timeout that triggers the re-broadcast of the packet\;
$next\_hop$ is the Link Layer address of the next hop\;
$broadcast$ is the Broadcast Link Layer address\;

\eIf{$p$ is newly generated by ego node}
{
  Set $next\_hop \gets broadcast$\;
  Return $next\_hop$\Comment*[r]{Immediate transmission by source router}}
{
  \eIf{$p$ in $buffer$}
  {
   \tcc{Contending}
   Remove $p$ from $buffer$\;
   Stop Timer\;
   Discard $p$\;
   Return -1\Comment*[r]{Indicates that packet is discarded}
  }
  {
   Add $p$ to $buffer$\Comment*[r]{New packet}
   Set $valid\_position \gets pos\_sender \textrm{ exists and } acc\_sender = TRUE$ \Comment*[r]{Indicates that sender position is valid}
   
   \eIf{$valid\_position$ and $acc\_ego = TRUE$}
   {
    Set $dist \gets DISTANCE(pos\_sender, pos\_ego)$\;
    Set $timeout \gets T_{CBF}(dist)$\Comment*[r]{Using equation~\ref{eq:TempCBF}}
   }
   {
    Set $timeout \gets T_{CBF-MAX}$\Comment*[r]{Use maximum timer if the distance cannot be calculated reliably}
   }
   Start Timer($timeout$)\;
   Return 0\Comment*[r]{Indicates that packet is buffered}
  }
}
\If{$Timer(timeout)$ expires}
{
    Fetch $p$ from $buffer$\;
    Set $next\_hop \gets broadcast$\;
    Return $next\_hop$\;
}
\end{algorithm}

\section{ETSI CBF problems and proposed solutions}
\label{subsec:contributions}

\subsection{Improving network efficiency by suppressing unnecessary retransmissions} 

\subsubsection{Adding Duplicate Packet Detection to ETSI CBF}
\label{subsec:dpd}

The ETSI GeoNetworking specification~\cite{etsiNewGeoNetworking} states that \textcolor{black}{DPD} must be done for any forwarding algorithm except for CBF \textcolor{black}{(Algorithm~\ref{alg:ETSI_recv})}. DPD is applied before entering the forwarding algorithm itself and, if a packet is a duplicate, the packet is discarded and no further processing is done, including that the packet is not delivered to upper layers. DPD is based on keeping a \textcolor{black}{Duplicate Packet List (DPL) for every node in the Location Table. The Location Table in a vehicle is a data structure that keeps information about other known GeoNetworking nodes, each one identified by its GeoNetworking address. The DPL stores the sequence numbers of each received packet originated in the respective node. The DPL can be implemented with a circular buffer, so packets are forgotten eventually, but this should happen when they are not longer travelling through the network.}

However, in ETSI CBF, the DPD mechanism is not employed and the solution solely relies on the CBF forwarding algorithm itself to deal with duplicate packets. This results in losing the two functions of DPD: duplicate packets can be delivered multiple times to upper layers when the vehicle is inside the area of interest, and can be forwarded several times. This occurs because the mechanism in the CBF forwarding algorithm to avoid duplicate packets is only effective in an ideal operation of the algorithm, as we describe next.  

\begin{figure}[bth!]
	\centering
	\includegraphics[width=\textwidth]{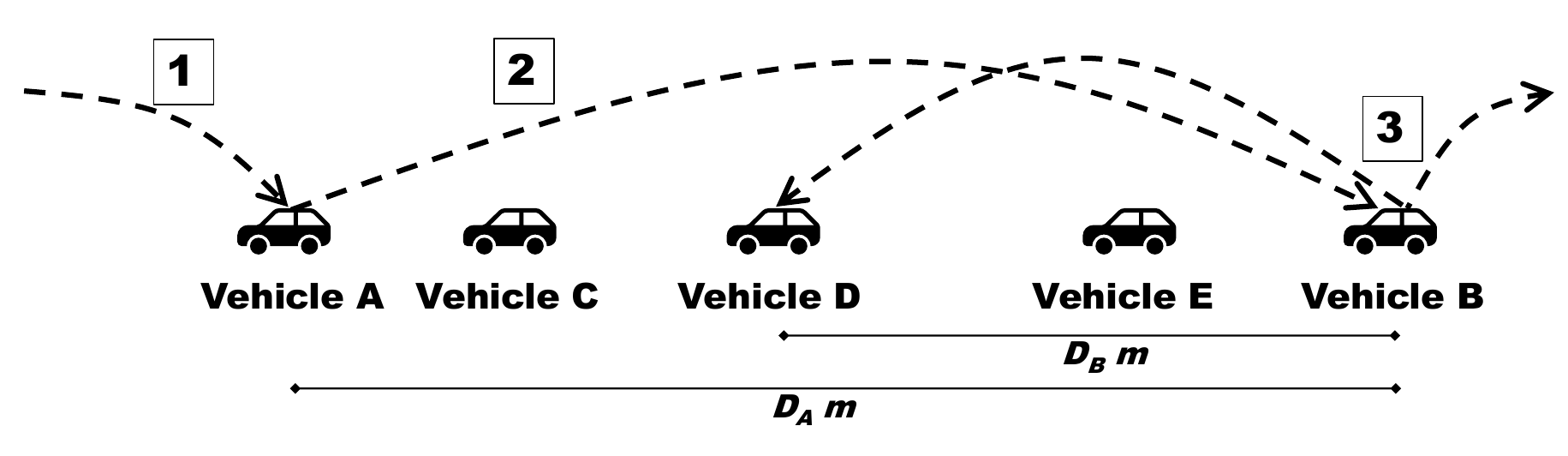}
	\caption{An example of a situation that generates duplicated packets with ETSI CBF forwarding}
	\label{fig:dpd_1}
\end{figure}

ETSI CBF, as explained in subsection~\ref{subsec:cbf}, checks each received packet against the packets in the CBF buffer (i.e., packets waiting to be forwarded). If a received packet is already in the CBF packet buffer, both, the received packet and the one in the buffer are discarded \textcolor{black}{(Algorithm~\ref{alg:ETSI})}, which has a similar effect to the DPD procedure. However, this procedure does not take into account the packets received in the past but not present in the CBF packet buffer anymore. This can happen because they are waiting to be transmitted in an underlying queue (e.g., DCC), or they have already been forwarded, or they have been discarded by a duplicate packet received previously. In an ideal world, this should never happen: when a vehicle sends a packet and the farthest forwarder (the one with the lowest contention timer value) transmits the same packet, all the vehicles that received the first one should also receive the second transmission and, thus, remove the packet from their CBF buffers simultaneously. However, in the real world, many common situations can lead to duplicate packets being received when the original packet is not longer in the CBF packet buffer. We describe some examples next:

\begin{itemize}
    \item When the source vehicle of a message sends it, it is transmitted directly, without being stored at its CBF buffer. Therefore, when another vehicle (which is in the radio coverage of the source) forwards the packet, the source vehicle will also receive the transmission and, since the packet is not in its CBF buffer, the source vehicle accepts its own packet as new, delivers it to the upper layer that generated it, and adds it to the CBF packet buffer for future retransmission.
    \item The situation we describe in this bullet is represented in Fig.~\ref{fig:dpd_1}. Vehicle A receives a message (point 1 in Fig.~\ref{fig:dpd_1}) or it could also be the message source. Some time later, because vehicle A is the best forwarder, it transmits the packet, which reaches all vehicles within a coverage of \textit{$D_A$}~meters (point 2 in Fig.~\ref{fig:dpd_1}). According to CBF rules, the farthest vehicle is the first to forward the packet. Suppose this second forwarder (vehicle B in the Fig.) is at nearly \textit{$D_A$}~m from the original forwarder (vehicle A) but, due to radio propagation issues or collisions, the radio coverage of vehicle B's transmission (point 3 in the Fig.) is \textit{$D_B < D_A$}~m. This means that some vehicles (e.g., vehicle C) close to the first forwarder (vehicle A) will not hear the second transmission (from vehicle B) and thus, unlike the vehicles within the \textit{$D_B$}~m coverage of vehicle B's transmission (e.g., vehicle E), they will not remove the previous packet received from vehicle A from their CBF buffers. Eventually, one of these vehicles (e.g., vehicle C)  will forward the packet they had received from vehicle A. Then, vehicles that have received both vehicle A and vehicle B transmissions (e.g., vehicle D and vehicle E), when receiving this new transmission of the packet from a new forwarder (e.g., vehicle C), will no longer have the packet in the CBF buffer (because it has been removed from there due to the transmission of vehicle B) and, therefore, they will add the packet again to the CBF buffer. Moreover, vehicles A and B will also add the packet to their respective CBF buffers, even when they have already forwarded the very same packet.
\end{itemize}

In general, any situation in which, for any reason, a vehicle located between two forwarders receives the transmission from one forwarder but not from the other (including source nodes) will result in the vehicle transmitting the packet when all of its neighbours have already removed it from their buffers. Hence, the neighbours will accept the packet again (delivering it to the upper layers) and will add the packet to their CBF buffers and thus, a second copy of the packet will be forwarded to progress through the area of interest. The end result is that, with ETSI CBF, there may be multiple waves of retransmissions of the same packet in the area of interest. This generates a very significant number of excessive retransmissions (that we will quantify in section~\ref{sec:evaluation}), causing an effect similar to a broadcast storm as the ones caused by simple flooding~\cite{tseng:2002}.

In order to solve these situations, we propose to  \textcolor{black}{include a DPD} mechanism also in the CBF algorithm specified by the ETSI GeoNetworking standard (note that the use of some kind of DPD mechanism with CBF has been proposed before, for example, in \cite{Kuhlmorgen2020, Paulin2015}). In our DPD mechanism, each vehicle stores the originating GeoNetworking address and sequence number of received packets in a \textcolor {black}{DPL} (as the rest of GeoNetworking algorithms do), plus a flag called \textit{new\_added} whose function will be explained next. Once a packet is received \textcolor{black}{(Algorithm~\ref{alg:DPD_recv})}, it is checked against the information in the \textcolor {black}{DPL}. If the packet is a duplicate (its information is already in the \textcolor {black}{DPL}), the packet is not delivered to upper layers, but it is still delivered to the CBF forwarding algorithm (to check if it is in the CBF buffer). On the other hand, if the packet information is not in the \textcolor {black}{DPL}, the packet is delivered to upper layers and to the CBF forwarding algorithm, and its information is added to the \textcolor {black}{DPL} with the flag \textit{new\_added} set. Once in the CBF forwarding algorithm \textcolor{black}{(Algorithm~\ref{alg:DPD})}, the packet is checked against the ones waiting in the CBF buffer. If the packet is not in the CBF buffer, then the \textcolor {black}{DPL} is checked again:
\begin{itemize}
\item If the packet is in the \textcolor {black}{DPL} with the \textit{new\_added} flag set, the flag is cleared, and the algorithm proceeds with its normal operation, adding it to the CBF packet buffer with the right timer for future retransmission.
\item If the packet is in the \textcolor {black}{DPL} with the \textit{new\_added} flag cleared (i.e. it was already added to the CBF buffer before), the packet is discarded. 
\end{itemize}
On the contrary, if the received packet is in the CBF buffer, it is discarded and the stored packet is also removed from the CBF buffer (as the current standard does). 

\begin{algorithm}[p] \footnotesize
\caption{\textcolor{black}{Duplicate Packet Detection (DPD) forwarding and receiver operations}}
\label{alg:DPD_recv}
\KwIn{$p = \textrm{the received GeoBroadcast packet}$}
$SO$ is the source of $p$\;
$pos\_SO$ is the position vector of $SO$\;
$pos\_ego$ is the position vector of the ego vehicle\;
$RHL$ is the Remaining Hop Limit of $p$\;
$SCF$ if the Store-Carry-Forward bit of the Traffic Class field of $p$\;
$LocT$ is the Location Table\;
$LS$ is the Location Service\;
$BCBuffer$ is the broadcast forwarding packet buffer\;

ProcessBasicHeader($p$)\;
ProcessCommonHeader($p$)\;
Set $F \gets InsideAreaOfInterest(pos\_ego)$\;
\If{$F < 0 \textrm{ and } itsGnNonAreaForwardingAlgorithm = 1~(GREEDY)$}
{
  \tcc{Outside the area of interest}
  \If{$p$ in $dpl$}
  {
    Discard $p$\;
    Return\;
  }
}
\If{$F >= 0$ and $(itsGnAreaForwardingAlgorithm = 1~(SIMPLE)$}
{
  \tcc{Inside the area of interest}
  \If{$p$ in $dpl$}
  {
    Discard $p$\;
    Return\;
  }
}
DuplicateAddressDetection($P$)\;
\eIf{$LocT(SO) \textrm{ does not exist}$}
{
    $LocT(SO) \gets pos\_SO$\;
    $LocT(SO).IS\_NEIGHBOUR \gets FALSE$\;
    UpdatePacketDataRate($LocT(SO)$)\;
}
{
    $LocT(SO) \gets pos\_SO$\;
    UpdatePacketDataRate($LocT(SO)$)\;
}
\If{$F >= 0$}
{
    \tcc{Inside the area of interest}
    \If{\textcolor{blue}{$p$ not in $dpl$}}
    {
        Indication($P.Payload$)\Comment*[r]{Pass the payload to upper protocol entity}
        \textcolor{blue}{Add $p$ to $dpl$}\;
        \textcolor{blue}{$dpl(p).new\_added \gets TRUE$};
    }
}
FlushPacketBuffers($SO$)\;
\eIf{$RHL <= 1$}
{
    Discard $p$\;
    Return\;
}
{
    $RHL \gets RHL - 1$
}
\If{$\forall LocT(SO).IS\_NEIGHBOUR = FALSE \in LocT \textrm{ and } SCF = TRUE$}
{
    \tcc{No neighbours exist}
    Add $p$ to $BCBuffer$\;
    Return\;
}
$ret \gets Forward(p)$
\If{$ret = 0~(Buffered)$ or $ret = -1~(Discarted)$}
{
    Return\;
}
MediaDependentProcedures($p$)\;
Transmit($p$)\;
\end{algorithm}

\begin{algorithm}[t!] \footnotesize
\caption{\textcolor{black}{Duplicate Packet Detection (DPD) forwarding algorithm}}
\label{alg:DPD}
\KwIn{$p = \textrm{the GeoNetworking packet to be forwarded}$}
$pos\_ego$ is the ego position vector with Position Accuracy Indicator $acc\_ego$\;
$pos\_sender$ is the sender position vector in the Location Table with Position Accuracy Indicator $acc\_sender$\;
$buffer$ is the CBF packet buffer\;
$timeout$ is the timeout that triggers the re-broadcast of the packet\;
$next\_hop$ is the Link Layer address of the next hop\;
$broadcast$ is the Broadcast Link Layer address\;
\textcolor{blue}{$dpl$ is the Duplicate Packet List\;}
\eIf{$p$ is newly generated by ego node}
{
  Set $next\_hop \gets broadcast$\;
  Set $timeout \gets T_{CBF-MAX}$ \;
  Return $next\_hop$ 
  }
{
  \eIf{$p$ in $buffer$}
  {
   Remove $p$ from $buffer$\;
   Stop Timer\;
   Discard $p$\;
   Return -1\;
  }
  {
  \eIf{\textcolor{blue}{$p$ in $dpl$ and $dpl(p).new\_added = FALSE$}}
   {
    \textcolor{blue}{Discard $p$\Comment*[r]{Ego node has received this packet previously}}
    \textcolor{blue}{Return -1}\;
   }
   {
       \textcolor{blue}{$dpl(p).new\_added \gets FALSE$}\;
       Add $p$ to $buffer$\;
       Set $valid\_position \gets pos\_sender \textrm{ exists and } acc\_sender = TRUE$\;
       
       \eIf{$valid\_position$ and $acc\_ego = TRUE$}
       {
        Set $dist \gets DISTANCE(pos\_sender, pos\_ego)$\;
        Set $timeout \gets T_{CBF}(dist)$ \;
       }
       {
        Set $timeout \gets T_{CBF-MAX}$ \;
       }
       Start Timer($timeout$)\;
       Return 0 \;
   }
  }
}

\If{$Timer(timeout)$ expires}
{
    Fetch $p$ from $buffer$\;
    Set $next\_hop \gets broadcast$\;
    Return $next\_hop$\;
}
\end{algorithm}

This way, if a vehicle does not cancel a packet from its CBF buffer (because it does not receive the transmission from the best forwarder, like Vehicle C in Fig.~\ref{fig:dpd_1}), it will still broadcast the packet to its neighbours. However, those neighbours will discard the duplicated packet because it is now listed in their respective \textcolor {black}{DPLs} (with a cleared \textit{new\_added} flag), thus effectively cancelling a second forwarding wave.

We also propose to add the packets originated in a vehicle (i.e., source node) to its \textcolor {black}{DPL} with the \textit{new\_added} flag cleared. This ensures that, if a vehicle receives a retransmission of its own packet from other vehicle, the source will neither deliver the message to the upper layers, nor it will execute the forwarding process.

\subsubsection{Suppression of Greedy Forwarding collisions at the area border}

Another source of undesired retransmissions may happen when forwarded packets inside the area of interest reach the area border. Vehicles outside the area of interest will receive packets broadcast from inside the area. To avoid further retransmissions, the standard has a clause that indicates that packets received from inside the area of interest should be discarded by vehicles outside the area. However, this mechanism does not always work as expected. The location of the vehicles at the border of the area of interest can be imprecise as seen by their neighbours, so a vehicle forwarding the packet from inside the area can be considered being outside the area by a neighbour. This can happen because the last location update or CAM sent by the forwarding vehicle occurred when the vehicle was still outside the area. When the non-area forwarding algorithm is the Greedy Forwarding algorithm (the default one for non-area forwarding in the ETSI GeoNetworking standard), all vehicles receiving a broadcast message from inside the area, but that think that it was sent from outside the area, will immediately try to send it back to the destination (towards the area of interest) using Greedy Forwarding. This may create retransmission storms at the area border and, since the Greedy Forwarding algorithm does not apply any kind of jitter to packets being forwarded, this could generate large collisions when multiple vehicles outside the area try to forward the same broadcast packet as soon as they receive it. If the DPD mechanism proposed above is not in use, the performance suffers even more, because the packet could be re-injected and forwarded again inside the area of interest.

We propose a simple mechanism to avoid these transmissions at area borders. If a vehicle is using the Greedy algorithm as non-area forwarding algorithm, the vehicle must never forward a packet received with the broadcast address as its link layer destination address. The motivation for this strategy is that, when a packet is progressing to the destination with the Greedy Forwarding algorithm, the destination link layer address must be a unicast MAC address, while the CBF algorithm used for forwarding inside the area of interest uses the broadcast MAC address.

This way, vehicles outside the interest area will never try to forward a CBF packet transmitted by a vehicle inside, even if they have stale location information and incorrectly think it also comes from outside the area, because the CBF packet will have the broadcast MAC address and thus will not be forwarded by the Greedy Forwarding algorithm.

\subsection{Improving reachability of DENM messages}
\label{subsec:improving-reachability}

\subsubsection{Enabling source retransmission}
\label{subsec:srcrtx}

The ETSI CBF algorithm provides a quite reliable forwarding process, because usually there are multiple potential forwarders for each packet. So, if any of them fail, for instance due to a collision with a close neighbour (notice that broadcast packets are not ACKed and thus cannot be retransmitted in case of collisions), the remaining ones will still try to forward the packet.

However this description applies to all hops in the forwarding path but the first one (i.e., the source of a packet). If the transmission of the source fails (e.g., due to a collision) and no vehicle receives it, nobody else is able to retransmit it again, other than the source. Usually DENM messages are sent several times by the Facilities service or it requests the network layer to retransmit them (usually with a spacing in the order of seconds), so the effect of losing one message depends on the requirements and settings specified for each scenario (e.g., ETSI specifies a number and frequency of repetitions for certain DENM causes, while leaving these parameters open for others). However, adding more reliability at the source is quite simple: just inserting any packet sent by a source vehicle to its CBF packet buffer with the maximum timer value (100\,ms), thus making the source to become a last-resort forwarder of its own packets. 

This way, if the first packet is lost, the source will not listen to any forwarders sending it and will retransmit it itself. Whereas, if the first transmission has really succeeded, the source will hear the forwarded packet and remove the duplicated one from its CBF packet buffer. Notice that this improvement can only been implemented after adding the \textcolor {black}{DPL} to CBF as proposed in subsection~\ref{subsec:dpd}, or it will worsen the problems highlighted on the previous section.

\subsubsection{Geographically-aware CBF Packet Cancellation}
\label{subsec:gpc}

Intuitively, since CBF timers are inversely proportional to the distance to the previous sender, the vehicle farthest away (but still within the area of interest) that receives a packet should be the first one to forward it. Thus, it makes sense to use that first transmission to cancel all other packets waiting in the CBF buffers of other vehicles, which, by following this intuition, should be nearer to the previous sender (i.e., with less progress), and thus worse forwarders. Moreover, it can also be assumed that most, or at least the vehicles that are closer to the new forwarder will receive its transmission, and thus cancel their packets.

\begin{figure}[H]
	\centering
	\includegraphics[width=\textwidth]{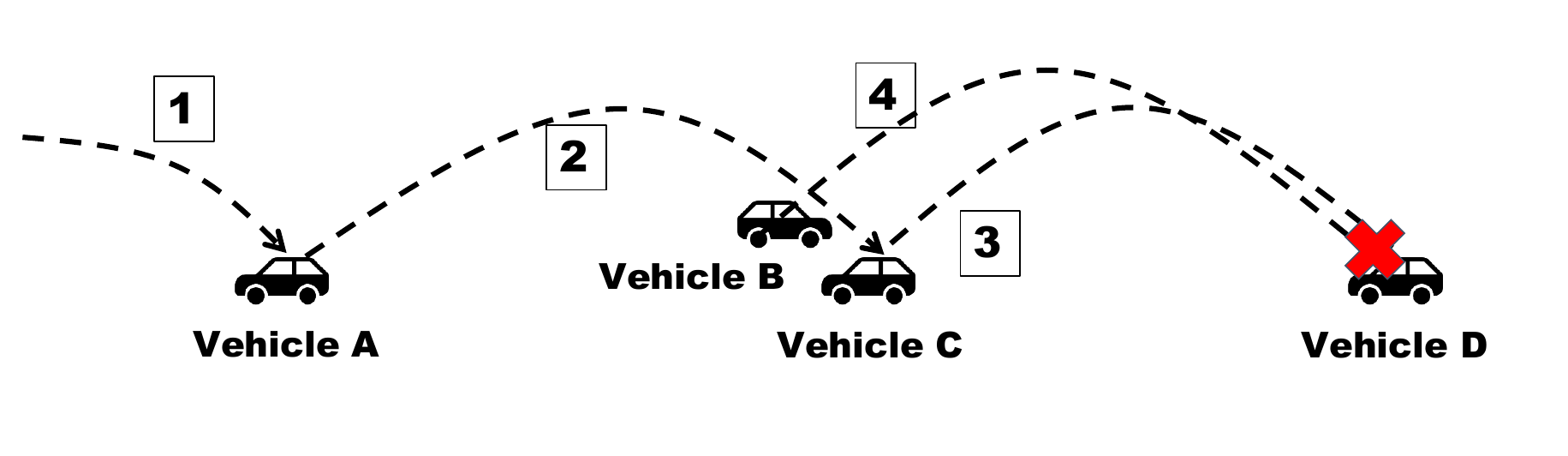}
	\caption{An example where a pair of vehicles aborts a message with standard CBF forwarding}
	\label{fig:gpc}
\end{figure}

However, the intuition underpinning this simple mechanism is not always true in real world scenarios. For instance, if two vehicles are running in parallel (Fig.~\ref{fig:gpc}) and thus are almost at the same distance from the sender, they will set their CBF timers with similar duration (e.g., that may expire less than 1\,ms away). Therefore, it may happen that when the first car transmits the packet, the second one has already sent its own to the lower layers, and thus it is not able to cancel its transmission. This leads to sending two times the same packet with a tiny time difference. Inconveniently, the effect of these pairs of packets is much worse that the obvious overhead, because the receivers will process the first one, storing it on the CBF buffer and setting a timer depending on its distance to the sender. But, immediately after, they receive the second packet that just cancels the first one before any of the timers of the potential forwarders can expire, effectively aborting the forwarding of the packet, unless there is some vehicle that receives just one of the packets (although, since both senders are close together, they will probably reach the same vehicles).

Actually, this situation may occur even if there are no vehicles running in parallel, due to the delay that the DCC mechanism may add to the packets being sent by the CBF layer (i.e., waiting in DCC queues). Let us image a row of vehicles that receive a packet from the sender. The timer of the farthest one ends first, but when the packet is sent to the lower layer to be transmitted, it has to wait an instant for the DCC gate to open (e.g., because it has just sent a CAM). Thus the timer of a second vehicle may also expire and, if this second packet does not have to wait at the DCC queue or just waits less that the first one, this second packet is transmitted before the first one, which cannot be cancelled, because it has already left the CBF buffer, leading to another pair of packets sent in close succession, and thus cancelling themselves mutually for all the next potential forwarders.

In order to solve these situations, we propose to modify the mechanism to cancel packets at the CBF buffer by taking into account the relative positions of the vehicle with respect to the sender of the duplicate packet, as well as the original source of the packet. The idea is to only cancel a packet in our CBF buffer with another one if we are sure that its sender is farther away from the original source than us, and they are not at different sides of the source vehicle. 

\begin{figure}[t]
	\centering
	\includegraphics[width=\textwidth]{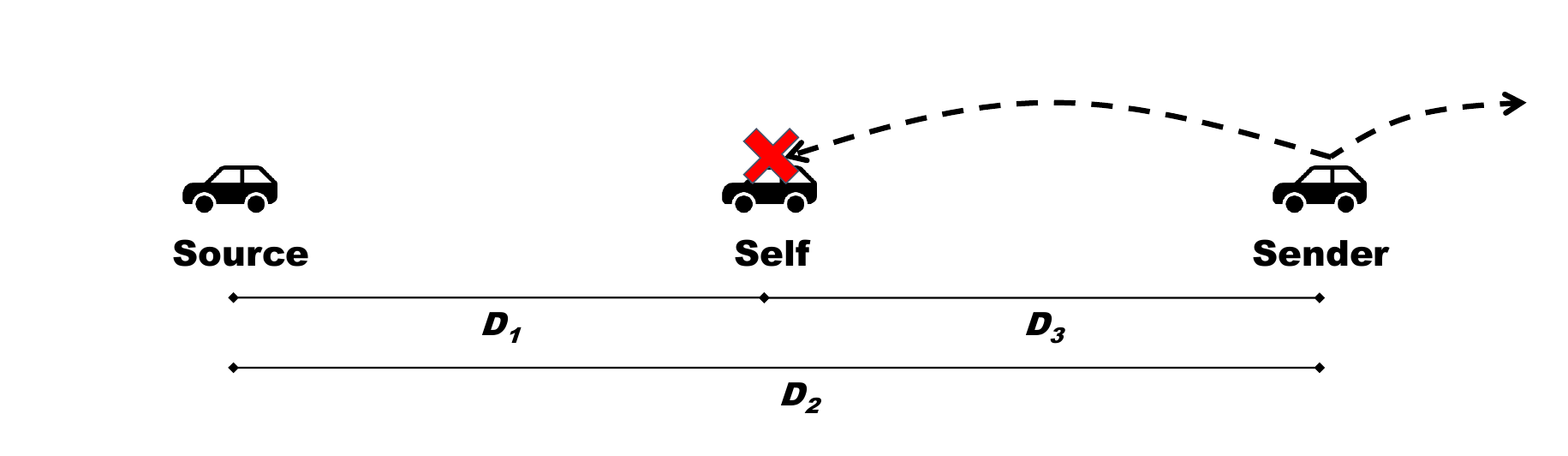}
	\caption{A vehicle cancelling a packet in its CBF Buffer with GPC because a better forwarded has transmitted it}
	\label{fig:gapc_1}
\end{figure}

In order to do so, the ego vehicle (Self in Fig.~\ref{fig:gapc_1}) that receives a packet (originally generated by Source vehicle in Fig.~\ref{fig:gapc_1}) sent from a neighbour (Sender in Fig.~\ref{fig:gapc_1}) has to compute the following distances:

$D_1$: the distance between the Source vehicle and the ego vehicle. The location of the Source vehicle is included in the packet, and each vehicle is always aware of its own location. 

$D_2$: the distance between the Sender and the Source vehicle. Since the Sender vehicle is a neighbour of the ego vehicle, its location can be obtained from the Location Table. If not present, we assume $D_2 = 0$ and thus the packet is not cancelled.

$D_3$: the distance between the Sender vehicle and the ego vehicle.

Then, when a packet is received in a vehicle and that packet is already present in the CBF buffer, the received packet is always dropped, but the packet at the CBF buffer is only cancelled if the following inequality occurs:

 \begin{equation}
\label{eq:gbc}
D_1 < D_2 \text{ and } D_2 > D_3 
\end{equation}

\begin{figure}[t!]
	\centering
	\includegraphics[width=\textwidth]{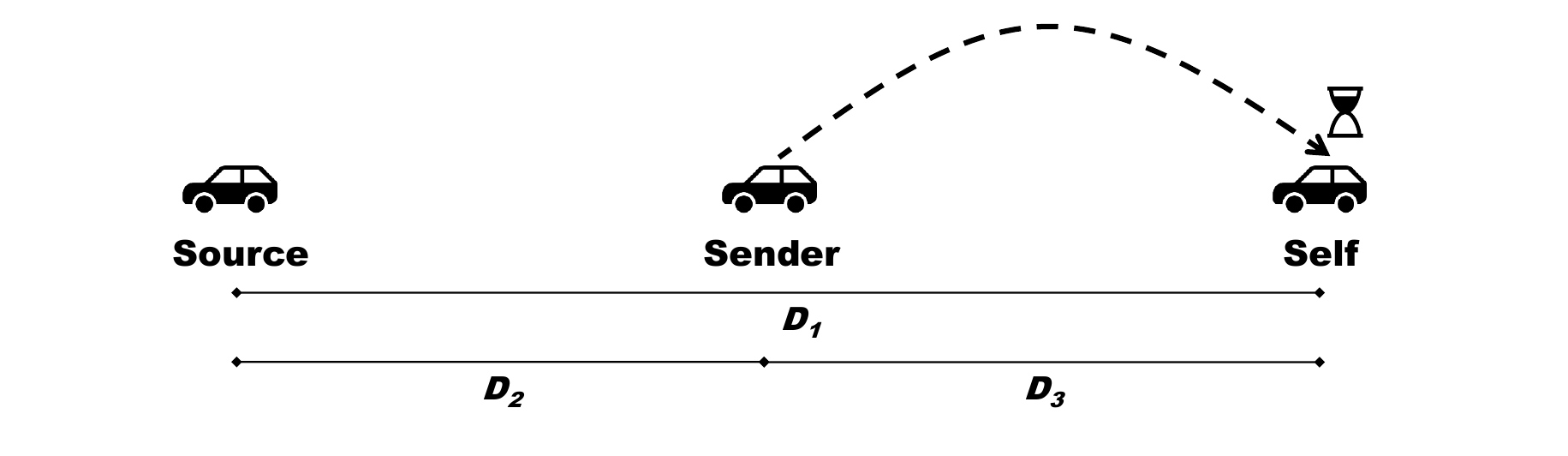}
	\caption{A vehicle reschedules, instead of cancelling, a packet in its CBF Buffer with GPC because a worse forwarder has just transmitted it}
	\label{fig:gapc_2}
\end{figure}

\begin{figure}[t!]
	\centering
	\includegraphics[width=\textwidth]{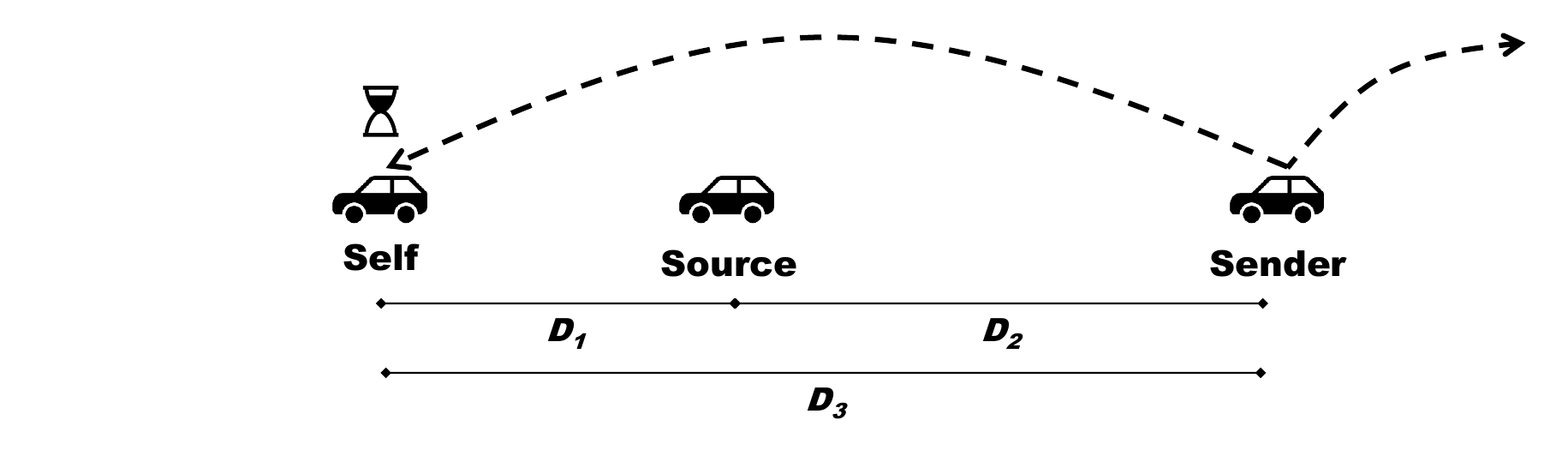}
	\caption{A vehicle reschedules, instead of cancelling, a packet in its CBF Buffer with GPC because a forwarder in other direction has just transmitted it}
	\label{fig:gapc_3}
\end{figure}

The \textcolor{black}{first inequality} is quite obvious: it checks that the Sender is actually farther away from the Source ($D_2$) than \textcolor{black}{Self} ($D_1$). Therefore in Fig.~\ref{fig:gapc_1} the Self vehicle has to cancel its CBF buffer packet, whereas in Fig.~\ref{fig:gapc_2} the Self vehicle will not. The second part requires a bit of explanation. If the Source vehicle broadcasts a message in an area of interest around it, it may happen that a Sender on one side of the Source incorrectly cancels all the forwarders on the other side, as illustrated by Fig.~\ref{fig:gapc_3}. Therefore, we can check that both the Sender and us are on the same side of the Source just by comparing $D_3$ and \textcolor{black}{$D_2$}.

When a vehicle receives a duplicated packet, but the above inequality does not hold (because the Sender of the packet is closer to the Source or at the other side of it) it discards the old packet in the buffer, but then inserts the received one in the CBF buffer, setting its timer according to the distance with the Sender, by following the standard CBF algorithm (i.e., the farther away from the Sender, the less time is waited). This rescheduling mechanism makes the vehicles close to the new Sender to wait more time, just in case the new transmission finds a forwarder farther away, that thus may be able to cancel these duplicated packets (this behaviour will be explained in more detail in the next section). But it is important to remember that the packets in the CBF buffer not fulfilling the inequality in Eq.~\ref{eq:gbc} are just delayed, not cancelled (unless a better forwarder does so later on). Thus, if no better forwarder is found with the duplicated transmission, the packet will keep advancing.

\begin{algorithm}[p] \footnotesize
\caption{\textcolor{black}{Geographically-aware CBF Packet Cancellation (GPC)  and source retransmission}}
\label{alg:GPC}
\KwIn{$p$ = \textrm{the GeoNetworking packet to be forwarded}}
$pos\_ego$ is the ego position vector with Position Accuracy Indicator $acc\_ego$\;
$pos\_sender$ is the sender position vector in the Location Table with Position Accuracy Indicator $acc\_sender$\;
\textcolor{blue}{$pos\_source$ is the source position vector obtained from the Long Position Vector field in the packet\;}
\textcolor{blue}{$d1$ is the distance between $pos\_ego$ and $pos\_source$\;}
\textcolor{blue}{$d2$ is the distance between $pos\_sender$ and $pos\_source$\;}
\textcolor{blue}{$d3$ is the distance between $pos\_ego$ and $pos\_sender$\;}
$buffer$ is the CBF packet buffer\;
$timeout$ is the timeout that triggers the re-broadcast of the packet\;
$next\_hop$ is the Link Layer address of the next hop\;
$broadcast$ is the Broadcast Link Layer address\;
$dpl$ is the Duplicate Packet List\;

\eIf{$p$ is newly generated by ego node}
{
  Set $next\_hop \gets broadcast$\;
  \textcolor{blue}{Add $p$ to $buffer$\Comment*[r]{Safeguard for a collision in the first hop}}
  Set $timeout \gets T_{CBF-MAX}$\;
  \textcolor{blue}{Add $p$ to $dpl$ \Comment*[r]{Ego node remembers its own packet}}
  \textcolor{blue}{$dpl(p).new\_added \gets FALSE$\;}
  Return $next\_hop$ 
}
{
  \eIf{$p$ in $buffer$}
  {
    \textcolor{blue}{Set $valid\_position \gets pos\_sender \textrm{ exists and } acc\_sender = TRUE$\;}
       
    \If{\textcolor{blue}{not $valid\_position$ or $acc\_ego = FALSE$}}
    {
        \textcolor{blue}{Set $d2 \gets 0$\;}
        \textcolor{blue}{Set $d3 \gets 0$\;}
    }
    
   \eIf{\textcolor{blue}{$d1<d2$ and $d2>d3$}}
   {
    \tcc{\textcolor{blue}{The packet comes from a more advanced forwarder}}
    Remove $p$ from $buffer$\;
    Stop Timer\;
    Discard $p$\;
    Return -1\;
   }
   {
    \tcc{\textcolor{blue}{The packet comes from a forwarder closer to the source}}
    \textcolor{blue}{Set $dist \gets d3$\;}
    \textcolor{blue}{Update $Timer(timeout) \gets T_{CBF}(dist)$}\;}
    \textcolor{blue}{Return 0 \Comment*[r]{Indicate that you kept p in buffer}
   }
   
  }
  {
  \eIf{$p$ in $dpl$ and $dpl(p).new\_added = FALSE$}
   {
    Discard $p$\;
    Return -1\;
   }
   {
       $dpl(p).new\_added \gets FALSE$\;
       Add $p$ to $buffer$\;
       Set $valid\_position \gets pos\_sender \textrm{ exists and } acc\_sender = TRUE$\;
       
       \eIf{$valid\_position$ and $acc\_ego = TRUE$}
       {
        Set $dist \gets DISTANCE(pos\_sender, pos\_ego)$\;
        Set $timeout \gets T_{CBF}(dist)$\;
       }
       {
        Set $timeout \gets T_{CBF-MAX}$ \;
       }
       Start $Timer(timeout)$\;
       Return 0\;
   }
  }
}
\If{$Timer(timeout)$ expires}
{
    Fetch $p$ from $buffer$\;
    Set $next\_hop \gets broadcast$\;
    Return $next\_hop$\;
}
\end{algorithm}

This Geographically-aware CBF Packet Cancellation (GPC) mechanism \textcolor{black}{(Algorithm~\ref{alg:GPC})} solves the problems generated by pairs of packets, because now the vehicles will not cancel the first packet with the second one, since they are farther away from the Source than the second Sender (i.e. $D_1 > D_2$).

\subsection{Reducing the number of DCC-induced retransmissions: Forward on Time (FoT)}
\label{subsec:fot}

A drawback of the GPC mechanism is that, in order to improve the packet reliability, it reduces the number of CBF packets that are cancelled. Therefore, it may lead to a larger number of packet transmissions, specially in high-load scenarios where the DCC mechanism disarranges the transmission order set up by the CBF timers.

\begin{figure}[t]
	\centering
	\includegraphics[width=\textwidth]{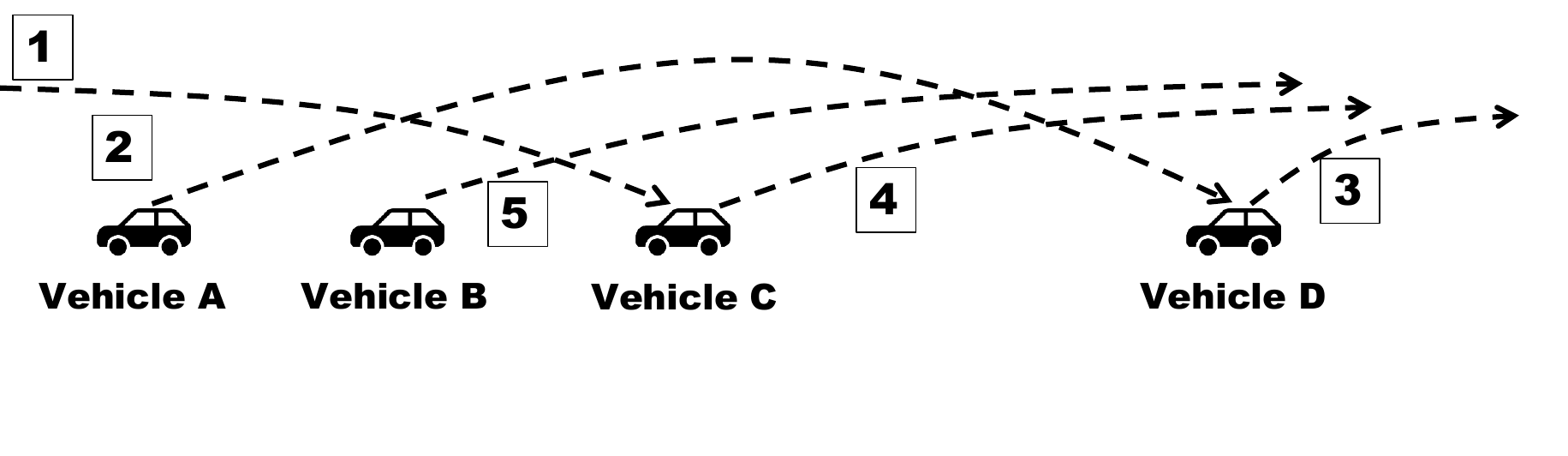}
	\caption{An example where DCC may induce several CBF packet transmissions}
	\label{fig:fot1}
\end{figure}

Let us illustrate this behaviour with an example (Fig.~\ref{fig:fot1}). Let us assume that vehicles A, B and C receive a packet (1) from a previous Sender or its Source and set up their CBF timers. Although vehicle C is the farthest one away, and thus has the shortest timer, let us assume that vehicles B and C cannot transmit their packets at the instant their CBF timers expire because they have just sent another packet (e.g., a CAM), and thus the forwarded DENM message must wait in the DCC queue until the DCC gatekeeper allows sending another packet. Therefore, vehicle A is the first one sending the packet (2). However, this packet cannot cancel the ones at vehicles B and C for two reasons: i) both vehicles B and C are farther away from the Source that vehicle A, so the GPC inequality does not hold, and thus they could only be delayed; \textcolor{black}{and ii)} in this example the packets of B and C are no longer in the CBF packet buffer at the network layer but waiting in a DCC queue, and hence cannot be cancelled by CBF mechanisms at all, even if they could receive later a transmission from a better forwarder (vehicle D) and the GPC inequality could now hold. Therefore, in this scenario the packet will be transmitted towards the next hop three times by vehicles A (2), B (5) and C (4). \textcolor{black}{The} only negative effect will be the transmission overhead, due to the previous mechanisms that will drop the duplicated packets (by using DPD), and because pairs of packets will not abort the forwarding (because of GPC).

K\"{u}hlmorgen et al.~\cite{Kuhlmorgen2020} have proposed solving this DCC-induced problem by a making DCC Access Layer check the Network Layer whenever a CBF packet is unqueued from a DCC queue, so DCC is able to drop duplicated CBF packets. Although this cross-layer mechanism indeed solves the above problem (e.g., the duplicated CBF packets from vehicles B and C could be cancelled, even if they have already left the CBF layer), this inter-layer communication is not considered by the current ETSI ITS architecture, and it requires a lower layer (DCC Access Layer) to be aware and understand the behaviour and messages of an upper one (CBF at the Network Layer), making both layers to be tightly coupled (notice that DCC provides service to all ETSI GeoNetworking algorithms, not only CBF). Besides, since \cite{Kuhlmorgen2020} does not include a GPC-like mechanism, this can result in wrong cancellations (in forwarders that are better than the Sender that is causing the cancellation).

As an alternative to \cite{Kuhlmorgen2020}, we propose to solve this problem with the so called Forward-on-Time (FoT) mechanism, which is a similar approach to Generate on Time (GoT)~\cite{got}. FoT only requires the CBF network layer to be aware of when the DCC gatekeeper will be open (this information can be shared through the DCC Cross-layer management plane), in order to keep the packets in the CBF buffer instead of at the DCC queues, and thus being able to reschedule and/or cancel them if necessary.

\begin{figure}[t]
	\centering
	\includegraphics[width=\textwidth]{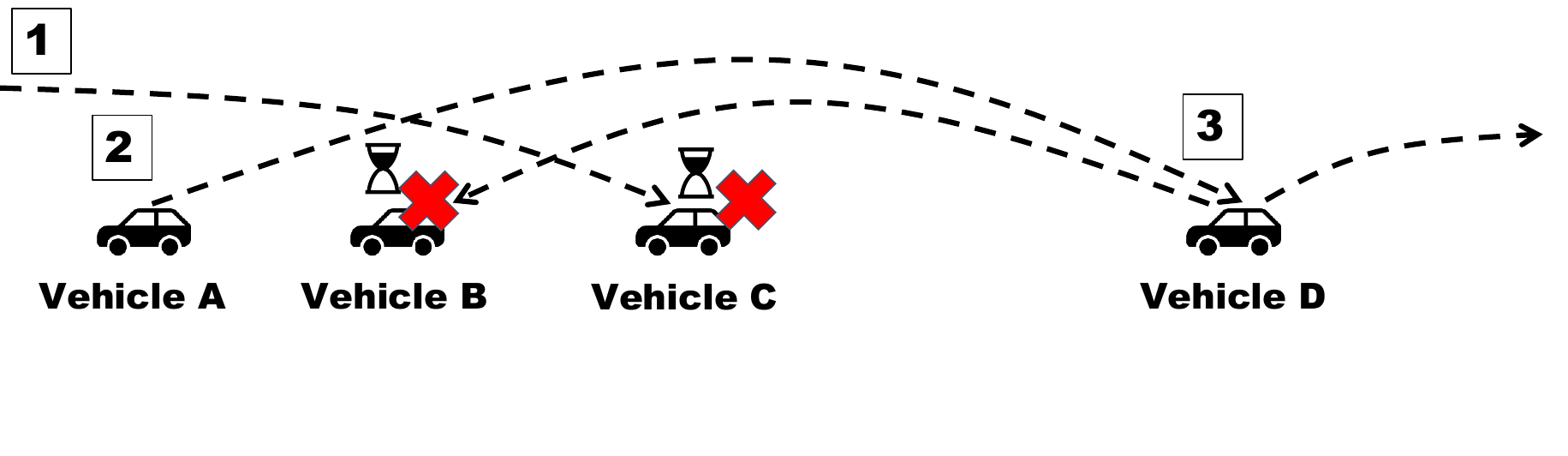}
	\caption{Reducing the number of DCC-induced CBF transmissions with FoT}
	\label{fig:fot2}
\end{figure}

\begin{algorithm}[p] \footnotesize
\caption{\textcolor{black}{Forward-on-Time (FoT)}}
\label{alg:FoT}
\KwIn{$p = \textrm{the GeoNetworking packet to be forwarded}$}
$pos\_ego$ is the ego position vector with Position Accuracy Indicator $acc\_ego$\;
$pos\_sender$ is the sender position vector in the Location Table with Position Accuracy Indicator $acc\_sender$\;
$pos\_source$ is the source position vector obtained from the Long Position Vector field in the packet\;
$d1$ is the distance between $pos\_ego$ and $pos\_source$\;
$d2$ is the distance between $pos\_sender$ and $pos\_source$\;
$d3$ is the distance between $pos\_ego$ and $pos\_sender$\;
$buffer$ is the CBF packet buffer\;
$timeout$ is the timeout that triggers the re-broadcast of the packet\;
\textcolor{blue}{$t_{go}$ is the time for the next transmission event at the DCC queue\;}
\textcolor{blue}{$t$ is the current time\;}
$next\_hop$ is the Link Layer address of the next hop\;
$broadcast$ is the Broadcast Link Layer address\;
$dpl$ is the Duplicate Packet List\;
\eIf{$p$ is newly generated by ego node}
{
  Set $next\_hop \gets broadcast$\;
  Add $p$ to $buffer$\;
  Set $timeout \gets T_{CBF-MAX}$\;
  Add $p$ to $dpl$\;
  $dpl(p).new\_added \gets FALSE$\;
  Return $next\_hop$\; 
  }
{
  \eIf{$p$ in $buffer$}
  {
    Set $valid\_position \gets pos\_sender \textrm{ exists and } acc\_sender = TRUE$\;
       
    \If{not $valid\_position$ or $acc\_ego = FALSE$}
    {
        Set $d2 \gets 0$\;
        Set $d3 \gets 0$\;
    }
  
   \eIf{$d1<d2$ and $d2>d3$}
   {
    \tcc{The packet comes from a more advanced forwarder}
    Remove $p$ from $buffer$\;
    Stop Timer\;
    Discard $p$\;
    Return -1\;
   }
   {
    \tcc{The packet comes from a forwarder closer to the source}
    Set $dist \gets d3$\;
    Update $Timer(timeout) \gets \textcolor{blue}{max(T_{CBF}(dist), t_{go}-t)}$\;
    Return 0\Comment*[r]{Indicate that you kept p in buffer}
   }
   
  }
  {
  \eIf{$p$ in $dpl$ and $dpl(p).new\_added = FALSE$\;}
   {
    Discard $p$\;
    Return -1\;
   }
   {
       $dpl(p).new\_added \gets FALSE$\;
       Add $p$ to $buffer$\;
       Set $valid\_position \gets pos\_sender \textrm{ exists and } acc\_sender = TRUE$\;
       
       \eIf{$valid\_position$ and $acc\_ego = TRUE$}
       {
        Set $dist \gets DISTANCE(pos\_sender, pos\_ego)$\;
        Set $timeout \gets \textcolor{blue}{max(T_{CBF}(dist), t_{go}-t)}$\;
       }
       {
        Set $timeout \gets \textcolor{blue}{max(T_{CBF-MAX}, t_{go}-t)}$\;
       }
       Start $Timer(timeout)$\;
       Return 0\;
   }
  }
}
\If{$Timer(timeout)$ expires}
{
    \eIf{\textcolor{blue}{$t \leq t_{go}$}}
    {
     \tcc{\textcolor{blue}{The DCC gatekeeper is open}}
     Fetch $p$ from $buffer$\;
     Set $next\_hop \gets broadcast$\;
     Return $next\_hop$\;}
    {
     \tcc{\textcolor{blue}{The DCC gatekeeper is closed}}
     \textcolor{blue}{Update $Timer(timeout) \gets t_{go} - t$\;}
     \textcolor{blue}{Return 0\Comment*[r]{Indicate that you kept p in buffer}}
    }
}
\end{algorithm}

In particular, the proposed Forward on Time (FoT) mechanism \textcolor{black}{(Algorithm~\ref{alg:FoT})} slightly modifies the set up of CBF packet timers by making them become the maximum value between the distance-based time computed by the standard ETSI CBF mechanism (Eq.~\ref{eq:TempCBF}), and the time left until the DCC gatekeeper opens. Therefore, if the CBF timer is greater than the DCC open period, it does not have any effect, but otherwise the packet is kept at the CBF buffer beyond its expiration time (instead of letting it go to wait in a DCC queue). Moreover, the DCC open period should also be checked whenever a CBF packet timer expires, since the DCC period may have been extended meanwhile (e.g., by sending a CAM while waiting for the CBF timer to expire). However, FoT cannot guarantee that forwarded CBF packets do not wait at DCC queues, since it may happen that, when the DCC gateway opens, a CAM is waiting at the DCC TC2 queue, and it is thus transmitted before the forwarded packet, which has to wait at the DCC TC3 queue until the DCC opens again (provided that there is no another higher priority packet waiting by then).

FoT solves the problem of DCC-induced packet disorder illustrated in the above example (Fig.~\ref{fig:fot2}), because with FoT the CBF packet timer of vehicles B and C will now be longer (since with FoT they also depend on the time until the DCC gatekeeper opens) than the one of vehicle A. This way, when A transmits its packet (2), vehicles B and C still have theirs in the CBF buffer and thus will now be able to reschedule it (in this case a large time because they are very close to vehicle A), thus letting another forwarder farther away (vehicle D) to retransmit it (3), and thus finally cancelling both.

Another benefit of FoT is that the ETSI GeoNetworking packet lifetime \cite{etsiNewGeoNetworking} better resembles the time it has been actually being forwarded, because the lifetime fields is only updated by the GeoNetworking layer (i.e. CBF buffering in this case). Therefore, it does not take into account the queuing time at layers below (especially DCC, which may be significantly larger than the CBF timer itself in high load scenarios).

\section{Evaluation}
\label{sec:evaluation}

\subsection{Simulation Scenarios}
\label{sec:sim_scenarios}

\textcolor{black}{For evaluation purposes we divide our proposals in three sets: 1) proposals to improve network efficiency by suppressing unnecessary retransmissions (i.e., adding Duplicate-Packet Detection and preventing Greedy Forwarding collisions at the area border), 2) proposals to improve the reachability of DENM messages (i.e., enabling source re-transmission, and implementing Geographically-aware CBF Packet Cancellation), and 3) proposals to reduce the number of DCC-induced retransmissions on CBF (i.e., implementing Forward-on-Time). To better understand the contribution to performance of each set of proposals, we add them incrementally to create three alternative mechanisms to the CBF algorithm specified by ETSI~\cite{etsiNewGeoNetworking}. For simplicity, we will refer to each evaluated mechanism as 'ETSI', 'DPD', 'GPC' and 'FoT'. ETSI is the ETSI CBF algorithm. DPD, GPC, and FoT implement the suppression of unnecessary retransmissions. GPC and FoT also implement the algorithms to improve the reachability of DENM messages, and, finally, FoT adds the Forward-on-Time algorithm. That is, each progressing mechanism includes the previous ones, and FoT includes all our proposals.}

\begin{figure}[h]
	\centering
	\includegraphics[width=\textwidth]{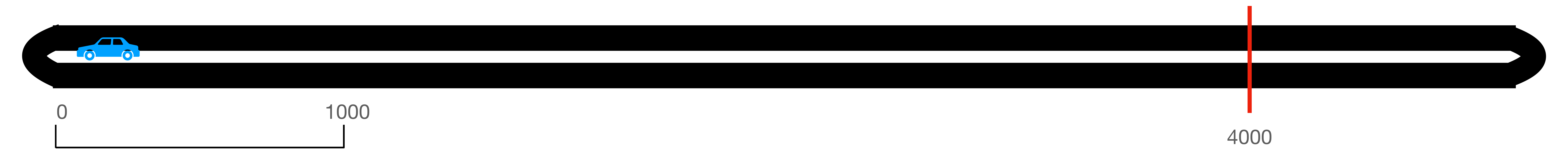}
	\caption{Highway scenario (vehicle size and lane widths are out of scale for better perception)}
	\label{fig:road}
\end{figure}

The four mechanisms are evaluated in a highway scenario: a 5 km straightaway in an 8-lane road (with four lanes per direction) where a stationary vehicle (e.g., due to a punctured tire) is located in a shoulder near the beginning of the stretch, as shown in Fig. \ref{fig:road}. It starts sending periodic warning messages (i.e., DENMs) at a frequency of 1 Hz generated at the Facilities level for 30 seconds. The area of interest is a rectangle that covers 4\,km of the straight stretch (all 8 lanes) behind the source vehicle and 100\,m in the front. This provides \textcolor{black}{a scenario} where messages have to hop several times to reach the border, while also allowing to observe the behavior of vehicles entering or departing the area of interest. Five traffic densities are measured: 10, 20, 30, 40, and 50 vehicles/km per lane, in order to measure the effect of network congestion.

\begin{table}[h]
	\centering
	\caption{Simulation Parameters}
	\label{tbl:simpars}
	\begin{tabular}{| l | l |}
		\hline
		\textbf{Parameter}  & \textbf{Values} \\
		\hline
		Access Layer protocol & ITS-G5 (IEEE 802.11p) \\
		Channel bandwidth & 10\,MHz at 5.9\,GHz \\
		Data rate & 6\,Mbit/s \\
		Transmit power & 20\,mW \\
		Path loss model & Simple Path-loss Model ($\alpha$ = 2.0) \\
		Maximum transmission range measured & 778\,m \\
		CAM packet size & 285 bytes \\
		CAM Traffic Class & TC2 \\
		DENM packet size & 301 bytes \\
		DENM Traffic Class & TC0 (Source) and TC3 (Forwarders) \\
		DENM lifetime & 10\,s\\
		\textcolor {black}{DPL} size & 32 packet identifiers per Source\\
		\hline
	\end{tabular}
\end{table}

Simulations are performed in Artery~\cite{Artery}, a simulation toolkit that implements ETSI ITS-G5. Artery has two main components: Artery itself, which runs on OMNET++ and is based on Veins~\cite{Veins}, and Vanetza --- an open-source C++ implementation of the ETSI C-ITS that comprises, among other protocols, the GeoNetworking and \textcolor{black}{DCC} mechanisms as specified by ETSI. To simulate vehicular traffic, Artery uses SUMO (Simulation of Urban MObility)~\cite{sumo2012}, and we use the default configuration for vehicle behavior (i.e., car-following, lane-changing, and speed models), and only passenger vehicles are included in the traffic flows. Parameters for the rest of the simulation, including those that affect GeoNetworking, are specified in Table \ref{tbl:simpars}.

The performance metrics that we evaluate are: number of transmissions of DENM messages (i.e., how many times a message is transmitted), Packet Delivery Ratio (i.e., the percentage of vehicles in the area of interest that receive the message), and end-to-end delay (i.e., how long it took the message to be received since it was generated). Measurements are taken for 35 seconds after a warming-up period that allows for vehicles to enter the simulation and integrate with the traffic flow. Statistics are collected from all vehicles inside the area of interest. Simulations are repeated 5 times with different random seeds, from which we obtain averages and confidence intervals (which are not shown in the figures because they are too small).

\subsection{Simulation Results}

\subsubsection{Number of transmissions}
As described in Section \ref{subsec:dpd}, the \textcolor{black}{CBF} algorithm as specified by ETSI only cancels the transmission of duplicate packets if they are received while another copy is still in its CBF buffer, but it does not keep a historical record of duplicate packets. This causes a phenomenon where even the source vehicle may receive and retransmit the message it has just released. Moreover, vehicles out of the area of interest may cause massive collision episodes at the border by wrongly forwarding CBF packets back to the area using Greedy Forwarding.

\begin{table}[h]
	\centering
	\caption{Average number of transmissions per density}
	\label{tbl:transmissions}
	\begin{tabular}{| c | r | r | r | r |}
		\hline
		\textbf{Density} & & & & \\
		\textbf{(veh/km lane)}  & \textbf{ETSI} & \textbf{DPD} & \textbf{GPC} & \textbf{FoT}\\
		\hline
		10 &  8,245.0 &   485.0 &   596.4 &   550.6 \\
		20 & 26,515.0 & 1,095.6 & 1,480.8 & 1,030.0 \\
		30 & 19,878.4 & 1,229.8 & 1,900.6 & 1,459.6 \\
		40 & 19,063.6 & 1,029.2 & 2,167.6 & 1,075.2 \\
		50 & 16,542.2 & 1,720.6 & 2,483.6 & 1,859.6 \\
		\hline
	\end{tabular}
\end{table}

Table~\ref{tbl:transmissions} shows that the presence of a Duplicate-packet Detection (DPD) mechanism and preventing the forwarding of broadcast packets by vehicles out of the area of interest causes transmissions to drop an order of magnitude (between 9.6 and 24.2 times depending on the traffic density). An analysis of simulation traces shows that, without DPD (i.e., the ETSI CBF standard), the source vehicle receives and forwards its own message several times, and a majority of vehicles in the system act as forwarders, even several times for the same packet. In fact, the traces show that many packets stop being forwarded only because of the default 10 hop limit parameter established by ETSI in its GeoNetworking Management Information Base (MIB)\cite{etsiNewGeoNetworking}. The analysis of simulation traces for the mechanisms including duplicate-packet detection (DPD, GPC, and FoT) shows that packets are now discarded because they have been found either in the CBF buffer or in the \textcolor {black}{DPL}. This effect is clearly observable in the number of transmissions of the DPD mechanism. GPC generates more transmissions than DPD, because it does not cancel every packet they find in the CBF buffer, only cancelling it if it comes from a forwarder that advances towards the border of the area. Finally, FoT makes packets wait longer at the CBF buffer instead of at DCC queues, and thus provides more opportunities for CBF packet suppression. This reduces the number of transmissions of FoT to a number similar to DPD (\textcolor{black}{reduction between 7.7\% and 50.5\% with respect to GPC}).

Table~\ref{tbl:transmissions} also shows the evolution in the number of transmissions when vehicle density increases. There is a significant jump from the 10 veh/km per lane to the 20 veh/km per lane densities, where the ETSI mechanism triples its transmissions, reaching its peak at least for the simulated densities. After 20 veh/km, the wireless channel begins to saturate, and DCC starts restricting the number of DENM messages that are able to make it into the medium, because CAM messages have a higher priority, thus slowly reducing the number of DENM transmission when vehicle density grows. \textcolor{black}{Similarly, DPD, GPC and FoT increase the number of transmissions  with vehicle density, up to a local maximum at around 30-40 veh/km for DPD and FoT (at a higher density than ETSI, since they transmit significantly less DENM packets) and then start decreasing until the largest density. Interestingly, for the largest density (50 veh/km per lane), we see that all DPD-enabled mechanisms show a peak in transmissions. This is due to DCC reaching a point where the time between allowed transmissions is significantly larger than CBF timers (max. 100\,ms), causing the CBF cancellation mechanism to become mostly ineffective.}

\subsubsection{Packet Delivery Ratio (PDR)}

The final objective of DENMs is to make as many vehicles within the area of interest aware of a hazardous situation. Thus, Packet Delivery Ratio (PDR) is a crucial metric to measure the success of a safety application. For this evaluation, we measure PDR as the ratio of vehicles that received a given message to the number of vehicles within the area at the moment the message was generated. Due to the fact that messages that are forwarded to extensive areas take some time to reach the border of the area of interest, PDR measurements also account for vehicles that depart or enter the area while the packet is being forwarded, and thus may rise above 100\%. We measure PDR at the Facilities Layer for the 30 DENM messages being sent during the simulation, and for the ETSI mechanism (which can receive the same message several times) we only consider the first message received by a station that is passed up from the Network Layer to the Facilities Layer.

\begin{table}[t]
	\centering
	\caption{Average Packet Delivery Ratio (PDR) per density}
	\label{tbl:pdr}
	\begin{tabular}{| c | c | c | c | c |}
		\hline
		\textbf{Density} & & & & \\
		\textbf{(veh/km lane)}  & \textbf{ETSI} & \textbf{DPD} & \textbf{GPC} & \textbf{FoT}\\
		\hline
		10 & 0.9998 & 0.9500 & 0.9917 & 0.9998 \\
		20 & 0.9961 & 0.9639 & 1.0036 & 1.0031 \\
		30 & 0.9280 & 0.8504 & 0.9864 & 1.0083 \\
		40 & 0.9371 & 0.8107 & 0.9852 & 1.0028 \\
		50 & 0.9372 & 0.8824 & 0.9812 & 0.9913 \\
		\hline
	\end{tabular}
\end{table}

Table \ref{tbl:pdr} shows the average PDR for all mechanisms for the five vehicle densities. For the lower densities (10 and 20), the ETSI mechanism achieves PDR values close to 100\%. This is due to the fact that vehicles are able to forward DENM messages freely, and its performance is due to massive flooding. For medium to high densities (30 to 50), the PDR achieved by the ETSI mechanism decreases. When observing Tables \ref{tbl:transmissions} and \ref{tbl:pdr}, we can see that PDR decreases along with the number of transmissions. This is due to the fact that forwarded messages (Traffic Class 3) have to coexist with CAM messages with higher priority (Traffic Class 2) thus, under high load, DCC prioritizes CAM messages over forwarded DENMs.

Results for the other three mechanisms can be divided in two sets. On the one hand, mechanisms with Geographically-Aware CBF Packet Cancellation (i.e., GPC and FoT), maintain PDR values close to 100\% across all densities. On the other hand, DPD reliability is 19\% lower than its GPC-enabled counterparts. This is explained by the phenomenon that is avoided by GPC --- forwarders could discard a packet because they receive another packet from a neighbour that is behind them. This can potentially cause scenarios where a message does not reach the border of the area, since all possible forwarders are prevented from advancing the message. Moreover, in some ETSI and DPD simulation traces no vehicle in the area of interest receives a given packet due to a collision when the source transmits it. This does no longer occur with GPC and FoT, where the source node inserts its packets on the CBF buffer and is able to retransmit them if the first transmission fails.

\begin{figure}[t]
	\centering
	\includegraphics[width=0.8\textwidth]{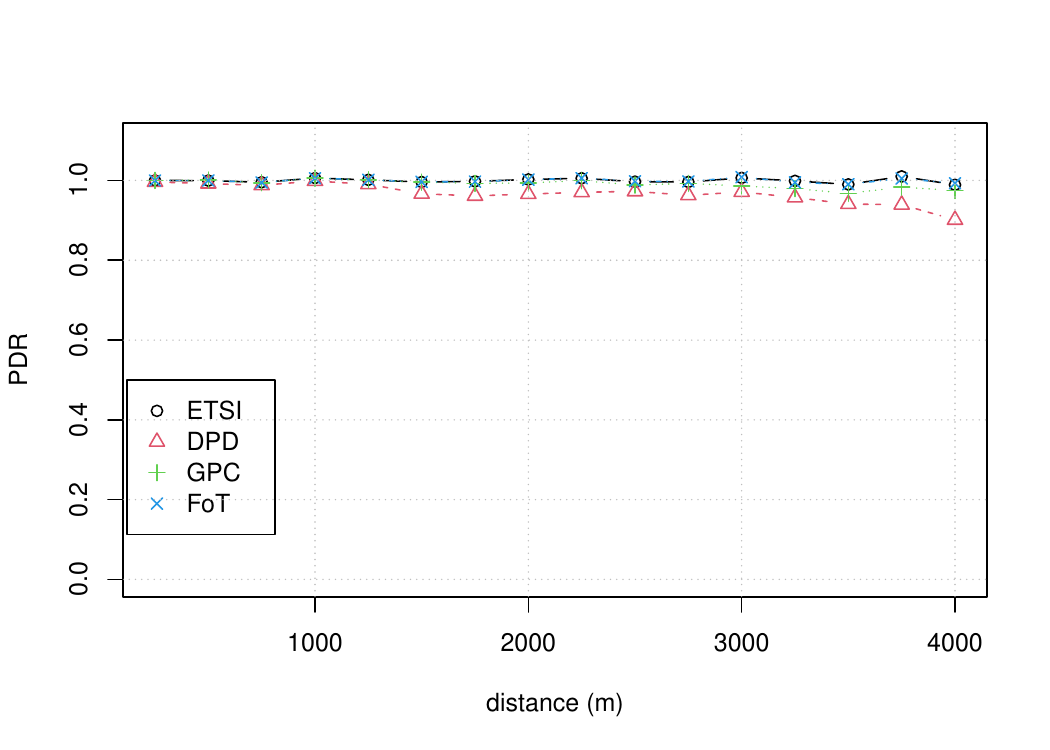}
	\caption{PDR for a density of 10\,veh/km per lane}
	\label{fig:pdr10}
\end{figure}

However, global PDR alone is not enough to assess reliability in multi-hop scenarios, since packet loss usually increases with the number of hops. Figure~\ref{fig:pdr10} shows the packet delivery ratio for DENMs aggregated by distance at the lowest density we simulate (10\,veh/km). All mechanisms achieve similar delivery ratios, close to 100\%. While the ETSI, GPC and FoT mechanisms obtain similar results, \textcolor{black}{GPC and FoT} achieve them while only transmitting a fraction of the messages that ETSI does. DPD, on the other hand, starts showing its shortcomings even at low densities: after one hop, it starts cancelling packets when they are forwarded by several vehicles that are close together (and thus have similar CBF timers).

\begin{figure}[th!]
	\centering
	\includegraphics[width=0.8\textwidth]{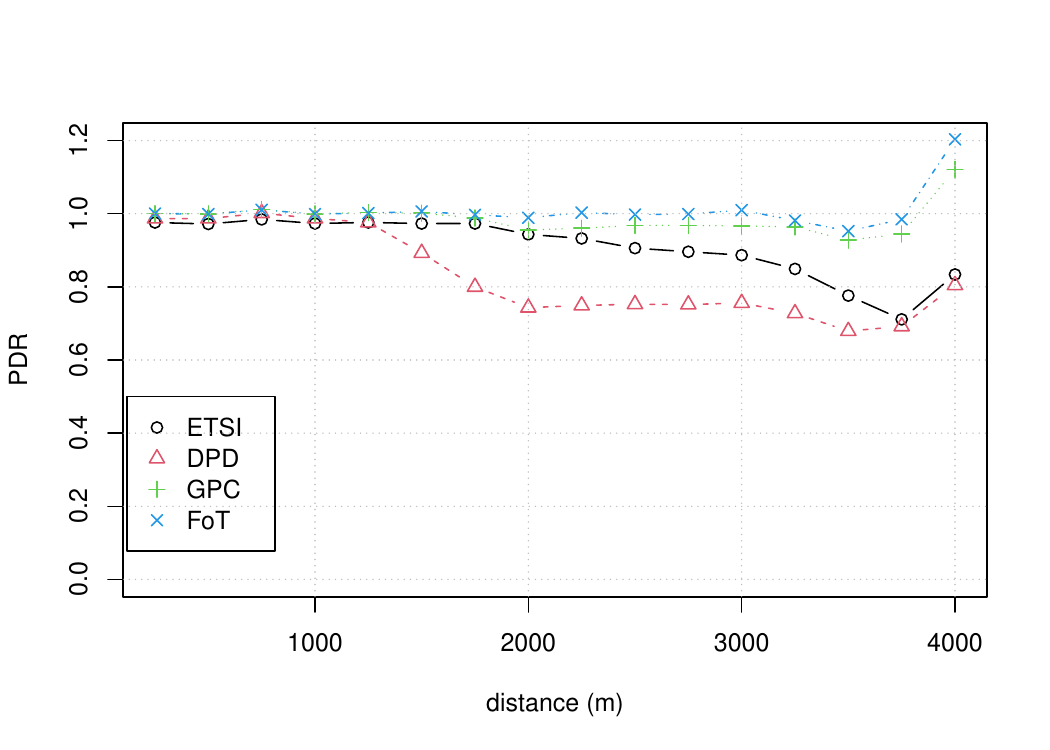}
	\caption{Packet Delivery Ratio (PDR) for a density of 30\,veh/km per lane}
	\label{fig:pdr30}
\end{figure}

Figure \ref{fig:pdr30} shows the packet delivery ratio for a density of 30\,veh/km per lane. At this medium density, \textcolor{black}{vehicles} face more restrictions by the cross-layer DCC mechanism: CAM generation is influenced by DCC as opposed to pure vehicle dynamics, generating messages at the allowed transmission rate. This causes forwarded DENMs to give way to higher-priority CAMs, which is reflected in the PDR behavior of ETSI and DPD, that are no longer able to achieve a 100\% delivery after one hop. ETSI outperforms DPD due to the sheer amount of messages that are transmitted, and massive flooding helps it increase the probability of messages being queued at DCC and finding a free slot in some vehicle to propagate to another forwarder. DPD continues to exhibit its reliability shortcomings, which are aggravated by DCC increasing the probability of a backlogged message cancelling all possible forwarders. On the other hand, GPC and FoT maintain a delivery ratio around 100\% throughout the whole area of interest.

Interestingly, all mechanisms exhibit a spike on the delivery rate in the last stretch of the area of interest. This is because vehicles that were not inside the area of interest when a packet was sent cross the border and receive the forwarded message as it becomes relevant to them.

\begin{figure}[t]
	\centering
	\includegraphics[width=0.8\textwidth]{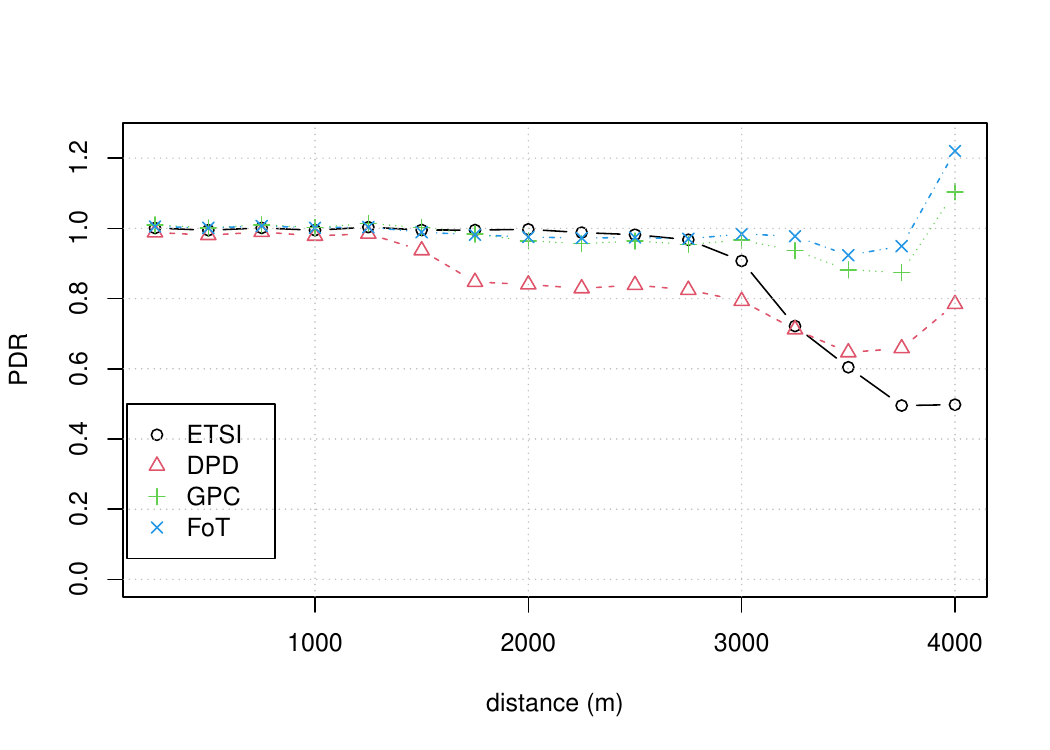}
	\caption{Packet Delivery Ratio (PDR) for a density of 50\,veh/km per lane}
	\label{fig:pdr50}
\end{figure}

Figure \ref{fig:pdr50} shows the delivery ratio for a density of 50\,veh/km per lane, which looks quite similar to results at 30\,veh/km. However at this high density, the time that DCC allows between DENM transmissions is higher than the maximum CBF timeout (100\,ms), thus rendering the CBF packet cancellation mechanism ineffective and thus leading to more transmissions in most mechanisms (although ETSI is so congested that DENM transmissions actually decrease). This leads to a slight reliability improvement for DPD, because these extra transmissions are able to recover the forwarding of packets that had been cancelled previously. Since GPC and FoT manage to maintain a delivery rate closer to 100\% along the whole area of interest, their reliability does not benefit from these extra transmissions.

\subsubsection{Latency}

We measure latency (i.e., end-to-end delay) for DENMs by \textcolor{black}{subtracting the time at which a particular message was generated at the Facilities layer at the source station from the time at which it is received at the Facilities Layer at a vehicle in the area of interest for the first time}. We obtain latency results from the full area of interest, including the first hop (transmitted with TC0), and the subsequent hops a message takes while attempting to reach the border (transmitted with TC3). We have aggregated the latency values of the 30 messages being sent across all 5 simulation repetitions.

\begin{figure}[h!]
     \centering
     \begin{subfigure}[h]{0.5\textwidth}
         \centering
	    \includegraphics[width=\textwidth]{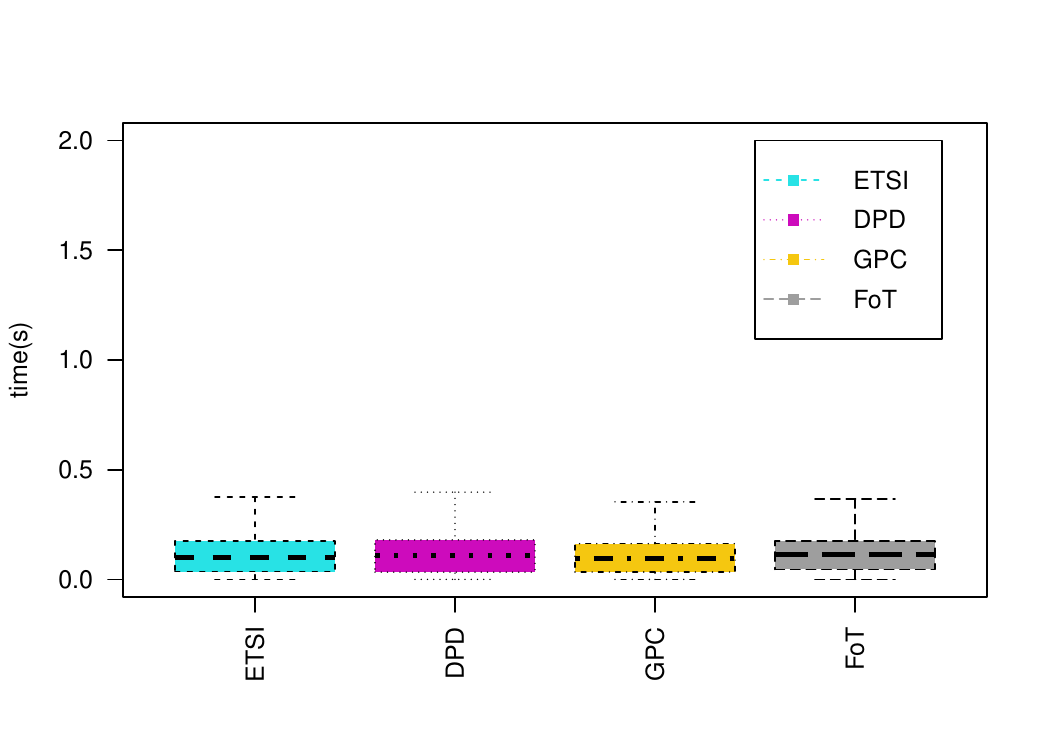}
	    \caption{Latency for a density of 10\,veh/km per lane}
	\end{subfigure}%
    ~
    \begin{subfigure}[h]{0.5\textwidth}
        \centering
	    \includegraphics[width=\textwidth]{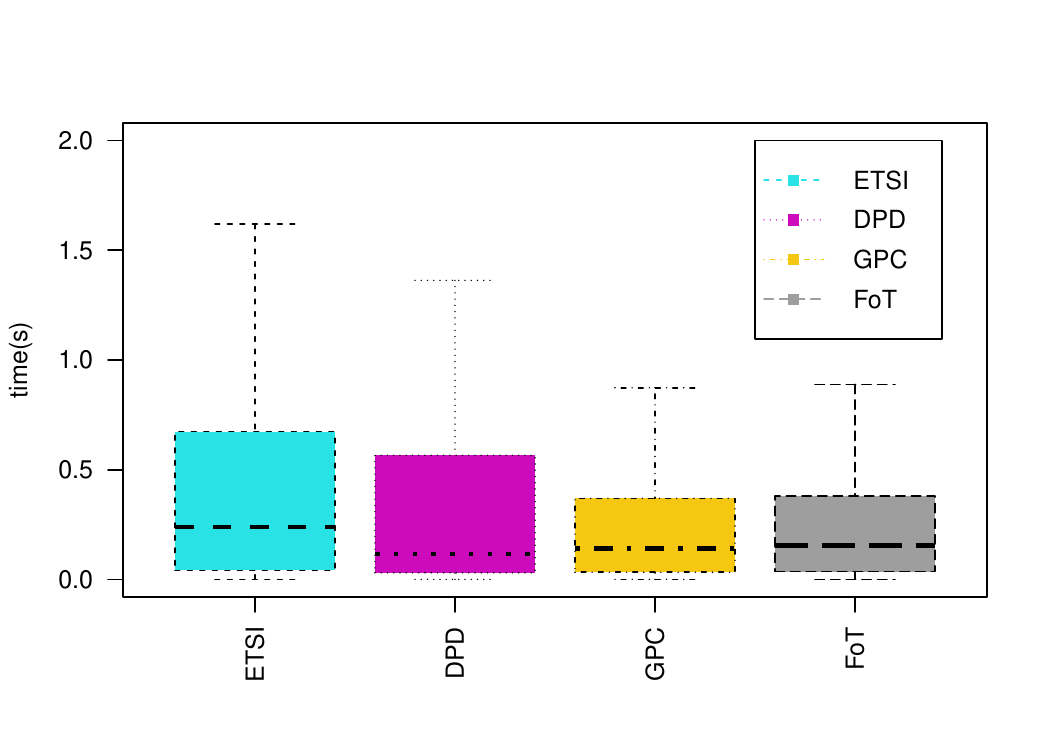}
	    \caption{Latency for a density of 30\,veh/km per lane}
	        \end{subfigure}
    \begin{subfigure}[h]{0.5\textwidth}
        \centering
	    \includegraphics[width=\textwidth]{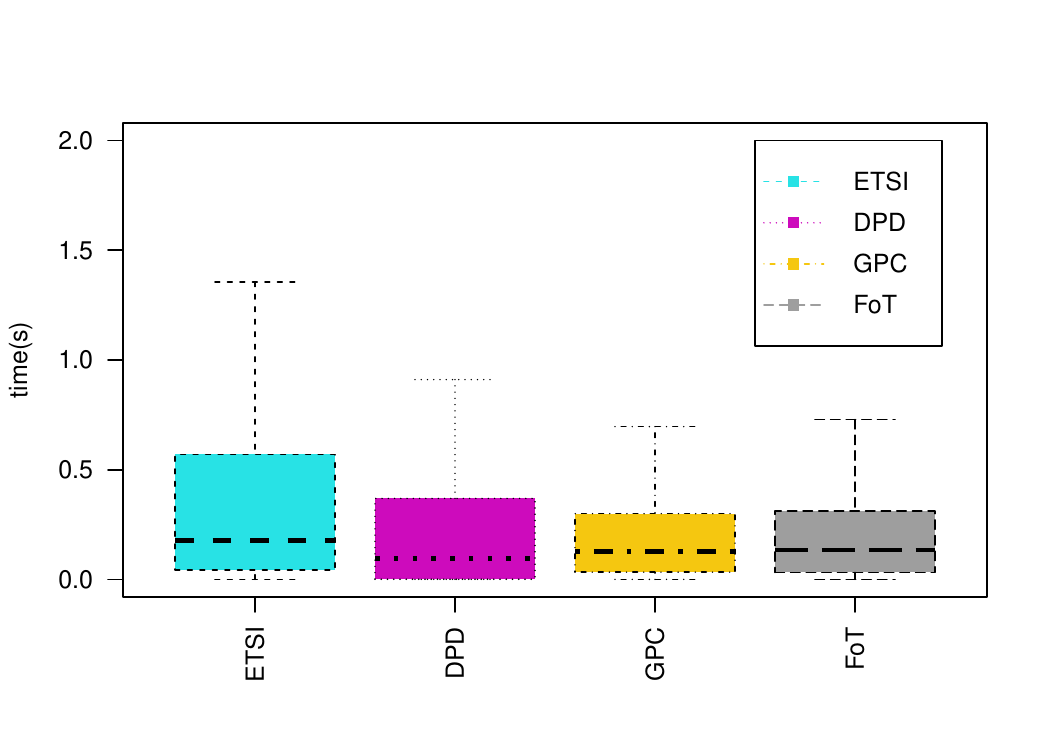}
	    \caption{Latency for a density of 50\,veh/km per lane}
	        \end{subfigure}
    \caption{Latency distribution for low, medium, and high densities}
    \label{fig:e2e}
\end{figure}

Fig. \ref{fig:e2e} shows the latency results for the 10, 30 and 50 veh/km per lane densities, excluding outlier values that would distort the time axis. Notice that due to vehicle dynamics a car can have a high CAM rate temporally, keeping a DENM message for several seconds at its DCC TC3 queue before being transmitted, and that packet can be received by any vehicle that entered the area of interest after the packet was generated. Results for the lowest density show that all mechanisms behave similarly, since at 10\,veh/km the PDR is almost 100\% for all mechanisms. However, once the network density increases, the number of losses due to collisions and packet cancellations also increases, leading to worse \textcolor{black}{latency} distributions, especially for ETSI and DPD mechanisms that do not implement GPC, and thus some packets can stop being forwarded but be retransmitted later. 

Although the \textcolor{black}{median latency in} all mechanism is similar at 30 and 50 veh/km densities, the small differences provide some interesting insights on the performance of the different mechanisms. For instance, DPD has surprisingly the lowest median latency of all mechanisms, but this is mainly an statistical artifact. Latency values are only gathered by vehicles that actually receive the packet. Therefore DPD, which has the worst PDR of all mechanisms, gathers less measurements from the last hops, and thus computing a smaller median value. However, ETSI, which also has a worse PDR than GPC and FoT, has the worst median \textcolor{black}{latency}. This result is because the back and forth waves of packets retransmissions that occur on ETSI CBF, where a packet that has stop being forwarded can be recovered by the retransmission of a packet that happens much later.

Finally, \textcolor{black}{it} is worth mentioning that not only do GPC and FoT achieve better reliability results, but also deliver the vast majority of DENM messages in less than 1 second across the 4 km area of interest, while the ETSI and DPD mechanisms exceed this threshold at high densities.

\section{Conclusions and Future Work}
\label{subsec:conclusions-fw}

\subsection{Conclusions}

After analyzing the standard ETSI Contention-Based Forwarding (CBF) algorithm \cite{etsiNewGeoNetworking} and its interaction with the Decentralized Congestion Control (DCC) \cite{etsiNewDcc} mechanism, we have identified the following problems with its performance and reliability:
\begin{enumerate}
\item The lack of long-term duplicate packet detection generates a massive number of transmissions, with waves of packet retransmissions going back and forth across the area of interest, that may make vehicles to receive and forward the same packet several times (even the source node that generated it), until it is finally dropped due to the hop limit.
\item A slight inaccuracy in neighbour positions may lead to massive congestion episodes at the border of the area of interest, when vehicles outside it may wrongly forward CBF packets back by using Greedy Forwarding.
\item The CBF algorithm is not reliable at the source node, because if the original transmission is lost (e.g., due to a collision), it is never retransmitted.
\item If two or an even number of vehicles transmit the same CBF packet consecutively (e.g., because they are in close proximity or induced by DCC delays), this could make all viable forwarders to cancel this packet from their CBF buffers, thus stopping its forwarding.
\item CBF packets can only be cancelled while they are stored at the CBF buffer, but in medium and high density scenarios they may wait at the DCC queue for a significant period of time, making the CBF cancelling mechanism less effective and leading to a high number of duplicate packet transmissions (that, on the other hand, compensate the previous cancellation problem).
\end{enumerate}

This paper has proposed a solution for each of these issues:
\begin{enumerate}
\item It is necessary to implement a Duplicate Packet Detection (DPD) mechanism in ETSI CBF in order to drop the duplicate packets that are no longer in the CBF buffer, so each CBF packet is only received and forwarded once.
\item The ETSI Greedy Forwarding algorithm should be modified in order to never forward a CBF packet back to the area of interest, by ignoring the packets that have been sent to the broadcast MAC address.
\item The source of a GeoBroadcast packet should also insert it in its CBF buffer, so it can be retransmitted if the first transmission fails (e.g., due to a collision).
\item The ETSI CBF packet cancellation algorithm should be modified to take into account the position of the sender with respect to the source and the ego vehicle, in order to decide whether the packet should be really cancelled (because it has been sent by a better forwarder) or merely being rescheduled (if the transmission has came from a worse one). The proposed algorithm is named Geographically-aware CBF Packet Cancellation (GPC).
\item A simple way to maintain the packets in the CBF buffer instead of in the DCC queue in order to increase the performance of the CBF packet cancellation mechanism is to compute the CBF timer taking also into account the time until the DCC gatekeeper allows the next transmission. The proposed improvement is called Forward-on-Time (FoT).
\end{enumerate}

We have evaluated all the proposed improvements and compared them to the standard ETSI CBF algorithm by means of simulation. Our results show that the implementation of a \textcolor {black}{DPD mechanism} in CBF and preventing forwarding CBF packets with Greedy Forwarding would reduce the total number of transmissions more than one order of magnitude (between 9.6 and 24.2 times depending on the vehicle density). Furthermore, by allowing the source node to be a last resort forwarded of its own packets, and replacing the ETSI CBF packet cancellation mechanism with the proposed \textcolor{black}{GPC} one, we are able to reach all vehicles (i.e., 100\% packet delivery ratio) inside the whole area of interest (an 8-lane, 4\,km highway stretch) in less than 1 second. Finally, implementing \textcolor{black}{FoT} to keep packets at the CBF buffer until DCC is able to transmit them reduces the number of transmissions of GPC between an 7.7\% and a 50.5\%.

\subsection{Future work}

Although the ETSI \textcolor{black}{CBF} algorithm is agnostic of the underlying access layer, some of the mechanisms that influence its behaviour, such as \textcolor{black}{DCC}, are only defined for the ITS-G5 access layer. Therefore it may be interesting to perform a similar evaluation for ETSI CBF running on top of C-V2X (cellular direct communications) access layer, in order to assess whether this stack has similar problems and the proposed solutions are also applicable.

The proposed \textcolor{black}{GPC} algorithm is remarkably simple to implement in linear topologies, such as the straight highway employed in the simulations. \textcolor{black}{We have also tested the proposed algorithms in a 2D scenario, where a circle-shaped area of interest encompasses two perpendicular highways, and the evaluation results are similar, reaching all vehicles inside the area of interest. Therefore,} we think it can also be successfully applied to the other area shapes supported by the ETSI GeoNetworking standard, \textcolor{black}{and more complex topologies. Although} this requires a detailed analysis and evaluation in other types of scenarios (e.g., urban ones).

Finally, we are currently studying how to improve the \textcolor{black}{FoT} mechanism to guarantee that forwarded packets do not end waiting at the DCC TC3 queue if \textcolor{black}{another} packet (e.g., a CAM) is already waiting at a higher priority queue (e.g., TC2).

\bibliography{etsi_cbf_evaluation}

\end{document}